\documentclass[english,prd,nofootinbib,twocolumn,floatfix,showpacs]{revtex4}
\usepackage{enumerate}
\usepackage{graphicx}
\usepackage{bm}
\usepackage{amsmath,amssymb}

\begin{document}
%------------------Local definitions--------------%
\def\PRL#1{Phys.\ Rev.\ Lett.\ {\bf#1}}\def\PR#1{Phys.\ Rev.\ {\bf#1}}
\def\ApJ#1{Astrophys.\ J.\ {\bf#1}}
\def\AaA#1{Astron.\ Astrophys.\ {\bf#1}}
\def\MNRAS#1{Mon.\ Not.\ R.\ Astr.\ Soc.\ {\bf#1}}
\def\CQG#1{Class.\ Quantum Grav.\ {\bf#1}}
\def\GRG#1{Gen.\ Relativ.\ Grav.\ {\bf#1}}
\def\IJMP#1{Int. J.\ Mod.\ Phys.\ {\bf#1}}
\def\JCAP#1{J.\ Cosmol.\ Astropart.\ Phys.\ #1}
\def\Z#1{_{\lower2pt\hbox{$\scriptstyle#1$}}} \def\w#1{\,\hbox{#1}}
\def\ns#1{_{\rm #1}} \def\Ns#1{_{\lower2pt\hbox{$\scriptstyle\rm#1$}}}
\def\ave#1{\langle{#1}\rangle} \def\srm#1{\hbox{$\scriptstyle\rm#1$}}
\def\lsim{\mathop{\hbox{${\lower3.8pt\hbox{$<$}}\atop{\raise0.2pt\hbox{$\sim$}}
$}}} \def\gsim{\mathop{\hbox{${\lower3.8pt\hbox{$>$}}\atop{\raise0.2pt\hbox{$
\sim$}}$}}} \def\kms{\w{km}\;\w{sec}^{-1}}\def\kmsMpc{\kms\w{Mpc}^{-1}}
\def\dd{{\rm d}} \def\ds{\dd s} \def\etal{{\em et al}}
\def\al{\alpha}\def\be{\beta}\def\ga{\gamma}\def\de{\delta}
\def\et{\eta}\def\th{\theta}\def\rh{\rho}\def\si{\sigma}
\def\ta{\tau} \def\tn{t\Z0}\def\taur{\tau\Ns{reion}}
\def\frn#1#2{{\textstyle{#1\over#2}}}\def\pt{\partial} \def\ab{{\bar a}}
\def\goesas{\mathop{\sim}\limits} \def\tw{\ta\ns{w}}
\def\gb{\bar\ga}\def\I{{\hbox{$\scriptscriptstyle I$}}}
\def\av{{a\ns{v}}} \def\aw{{a\ns{w}}}\def\Vav{{\cal V}}
\def\DD{{\cal D}}\def\gd{{{}^3\!g}}\def\Rav{\ave{\cal R}}
\def\QQ{{\cal Q}}\def\dsp{\displaystyle}
\def\mean#1{{\vphantom{\tilde#1}\bar#1}}\def\bx{{\mathbf x}}
\def\bH{\mean H}\def\Hb{\bH\Z{\!0}}\def\gb{\mean\ga}
\def\bD{\mean D}\def\rhb{\mean\rh}\def\OM{\mean\Omega}\def\eb{\mean\et}
\def\fw{{f\ns w}}\def\fv{{f\ns v}} \def\goesas{\mathop{\sim}\limits}
\def\fvn{f\ns{v0}} \def\fvf{\left(1-\fv\right)}
\def\OMM{\OM\Z M}\def\OMk{\OM\Z k}\def\OMQ{\OM\Z{\QQ}}
\def\dA{d_{\lower1pt\hbox{$\scriptstyle A$}}}
\def\da{d_{\lower1pt\hbox{$\scriptscriptstyle A$}}}
\def\hOmM{{\hat\Omega}\Ns{M}}\def\hOmR{{\hat\Omega}\Ns{R}}
\def\hOmk{{\hat\Omega}\Ns{k}}\def\hOmL{{\hat\Omega}\Z{\LA}}
\def\OmMn{\Omega\Ns{M0}}\def\OMBn{\OM\Ns{B0}}\def\OMCn{\OM\Ns{C0}}
\def\OmRn{\Omega\Ns{R0}}\def\Omgn{\Omega\Z{\ga\srm{0}}}
\def\Omkn{\Omega\Ns{k0}}\def\OmLn{\Omega\Z{\LA\srm{0}}}
\def\OMMn{\OM\Ns{M0}}\def\OMkn{\OM\Ns{k0}}
\def\OmBn{\Omega\Ns{B0}}\def\OmCn{\Omega\Ns{C0}}\def\OMRn{\OM\Ns{R0}}
\def\la{\lambda}\def\Cl{C_\ell}
\def\Hn{H\Z0}\def\fvi{{f\ns{vi}}}\def\fwi{{f\ns{wi}}}
\def\Hv{H\ns v}\def\Hw{H\ns w}\def\kv{k\ns v}\def\rb{\mean r}
\def\LA{\Lambda}\def\LCDM{$\LA$CDM}\def\obs{^{\rm obs}}
\def\zb{\bar{z}}\def\Tb{\mean T}\def\siT{\si\Ns{T}}
\def\rhM{\rh\Ns{M}}\def\rhR{\rh\Ns{R}}\def\pR{P\Ns{R}}\def\Tbn{\Tb\Ns{0}}
\def\OMRn{\OM\Z{R0}}\def\OMR{\OM\Z R}\def\abn{\ab\Ns{0}}\def\gbn{\gb\Ns{0}}
\def\h{\,h^{-1}}\def\hm{\h\hbox{Mpc}}\def\etBg{\et\Z{B\ga}}
\def\DE{\Delta}\def\Nn{N_\nu} \def\tl{\tilde\ell}\def\vt{\vartheta}
\def\tdec{t\ns{dec}}\def\hdec{\bar h\ns{dec}}\def\bxd{{\bar x}\ns{dec}}
\def\hOmMn{{\hat\Omega}\Ns{M0}}\def\hOmRn{{\hat\Omega}\Ns{R0}}
\def\hOmBn{{\hat\Omega}\Ns{B0}}\def\hOmkn{{\hat\Omega}\Ns{k0}}
\def\hOmLn{{\hat\Omega}\Z{\LA\srm{0}}}
\def\siB{\si\Ns{B}}\def\Neff{N\Ns{eff}}\def\dec{\,\rm dec}
\def\AH{A($\bH\ns{dec}$)}\def\Ar{A($\rb\Z{\cal H}$)}
\def\Ae{A($\eb\Z0$)}\def\At{A($\tn$)}\def\Ao{A($\hOmLn=0$)}
\def\Ww{W($k=0$)}\def\Wv{W($k\ne0$)}\def\OMd{\OM\Ns{dec}}
\def\OMMd{\OM\Z{M\dec}}\def\lOMk{\tOM\Z{k}}\def\Te{\widetilde\et}
\def\OMRd{\OM\Z{R\dec}}\def\OMkd{\OM\Z{k\dec}}\def\OMQd{\OM\Z{{\QQ}\dec}}
\def\tOM{\widetilde\Omega}\def\lOMM{\tOM\Z{M}}\def\lOMR{\tOM\Z{R}}
\def\TH{\widetilde H}\def\Ta{\tilde a}
\def\lOMMn{\tOM\Z{M0}}\def\lOMRn{\tOM\Z{R0}} \def\at{{\tilde a}}
\def\lOMkn{\tOM\Z{k0}} \def\atn{\at\Z0}
\def\lOMMd{\tOM\Z{M\dec}}\def\lOMRd{\tOM\Z{R\dec}}\def\Trh{\tilde\rh}
\def\lOMkd{\tOM\Z{k\dec}}
\def\Txd{\widetilde x\ns{dec}} \def\pAH{\protect\AH}
\def\beq{\begin{equation}} \def\eeq{\end{equation}}
\def\bea{\begin{eqnarray}} \def\eea{\end{eqnarray}}
%-------------------------------------------------%
%
\title{Cosmic microwave background anisotropies in the timescape cosmology}

\author{M.~Ahsan Nazer}
\email{Ahsan.Nazer@pg.canterbury.ac.nz}
\author{David~L.~Wiltshire}
\email{David.Wiltshire@canterbury.ac.nz}
\affiliation{Department of Physics \& Astronomy,
University of Canterbury, Private Bag 4800, Christchurch 8140, New Zealand}

\begin{abstract}
We analyze the spectrum of cosmic microwave background (CMB) anisotropies in
the timescape cosmology: a potentially viable alternative to homogeneous
isotropic cosmologies without dark energy. We exploit the fact that the
timescape cosmology is extremely close to the standard cosmology at early
epochs to adapt existing numerical codes to produce CMB
anisotropy spectra, and to match these as closely as possible to the timescape
expansion history. A variety of matching methods are studied
and compared. We perform Markov chain Monte Carlo analyses on the parameter
space, and fit CMB multipoles $50\le\ell\le2500$ to the Planck satellite
data. Parameter fits include a dressed Hubble constant,
$\Hn=61.0\kmsMpc$ ($\pm1.3$\% stat) ($\pm8$\% sys), and a present void
volume fraction $\fvn=0.627$ ($\pm2.3$\% stat) ($\pm13$\% sys). We find best
fit likelihoods which are comparable to that of the best fit \LCDM\ cosmology
in the same multipole range. In contrast to
earlier results, the parameter constraints afforded by this analysis no
longer admit the possibility of a solution to the primordial lithium
abundance anomaly. This issue is related to a strong constraint between the
ratio of baryonic to nonbaryonic dark matter and the ratio of heights of the
second and third acoustic peaks, which cannot be changed as long as
the standard cosmology is assumed up to the surface of last scattering. These
conclusions may change if backreaction terms are also included in the
radiation--dominated primordial plasma.
\end{abstract}
\pacs{98.80.-k 98.80.Es 04.20.-q.\qquad {\em Physical Review D} {\bf91}, 063519
(2015).}
\maketitle

\section{Introduction}

The timescape cosmology \cite{clocks,sol,obs} is a viable alternative to the
standard cosmology, without dark energy. It abandons the assumption that
average cosmic evolution is identical to that of an exactly homogeneous
isotropic Friedmann--Lema\^{\i}tre--Robertson--Walker (FLRW) model, but
relies on a simplifying principle concerning gravitational energy and
average cosmic evolution \cite{equiv} to obtain a phenomenologically predictive
model \cite{obs,lnw,sw,grb,dnw}.

The assumption of spatial homogeneity is observationally well justified at the
epoch of last scattering, given the evidence of the cosmic microwave background
(CMB) radiation. However, eventually the growth of structure becomes nonlinear
with respect to perturbation theory about a FLRW model, and by the present epoch
the Universe is only homogeneous in some average statistical sense when one
averages on scales $\gsim100\hm$, where $h$ is the dimensionless parameter
related to the Hubble constant by $\Hn=100h\kmsMpc$.

The exact spatial scale of the transition to average homogeneity, which we will
call the statistical homogeneity scale, is debated \cite{h05,sl09,sdb12}.
However, the most conservative estimates are that it occurs in the range
$70$--$100\hm$ \cite{sdb12}. This is consistent with the idea adopted in the
timescape scenario that it is the small amplifications of initial perturbations
by acoustic waves in the primordial plasma on scales smaller than the
$\goesas100\hm$ baryon acoustic oscillation scale which determine the
transition to the nonlinear regime \cite{bscg}.
Below the statistical homogeneity scale we observe a universe dominated in
volume by voids \cite{HV1,HV2,Pan11}, with clusters of galaxies in walls,
sheets and filaments surrounding and threading the voids. Voids
of diameter $\goesas30\hm$ form some 40\% of the volume of the present epoch
Universe \cite{HV1,HV2}. They represent the largest typical structures,
though there are some larger less typical structures \cite{Pan11,lts,nh14}, and
also a numerous population of smaller minivoids \cite{minivoids}.

The problem of fitting a smooth geometry \cite{fit1,fit2}
to the complex hierarchy of the
observed cosmic web requires addressing a number of fundamental questions
\cite{dust}, including: (i) how is average cosmic evolution to be described; and
(ii) how are local observables related to quantities defined with respect to
some average geometry? The fitting problem is not directly addressed in the
standard cosmology. One simply assumes that average evolution follows a FLRW
model. However, FLRW models expand rigidly to maintain constant spatial
curvature, a feature which is not generic in general relativity and which is
absent even in simple inhomogeneous models such as the
Lema\^{\i}tre--Tolman--Bondi (LTB) \cite{L,T,B} and Szekeres \cite{Sz}
solutions.

General inhomogeneous cosmological models which begin close to a spatially
homogeneous isotropic model will exhibit backreaction (see, e.g,
\cite{BR,CELU,vdH}), leading to a non-FLRW expansion. However,
the general situation is so complex that even when studying
inhomogeneity and backreaction, many researchers simply assume an average
FLRW expansion, given the phenomenological success of the standard model.
If ``dark energy'' is to have a deep physical explanation, however, then
the phenomenological success of the standard model should be taken as a signal
that there are simplifying principles to be found in the averaging problem.
The timescape model relies on such a simplification \cite{equiv}: there is
always a choice of clocks and rulers of canonical observers which makes the
average Hubble expansion uniform despite inhomogeneity.

To a achieve a notion of average uniform but nonrigid expansion one gives up
the assumption that in the presence of spatial curvature gradients the clocks
of all average observers are synchronized.
It is well understood that observers at fixed spatial coordinates in a standard
perturbed FLRW universe with geometry,
\beq
\ds^2=-(1+2\Phi)c^2\dd t^2+a^2(1-2\Psi)\de_{ij}\dd x^i\dd x^j,\label{FLN}
\eeq
in Newtonian gauge, will have clock rates that differ slightly from cosmic
time, $t$, on account of their position in the gravitational potential
$\Phi(t,\bx)$.
However, although all actual observers exist in bound structures deep within the
nonlinear regime of perturbation theory, a time dilation effect of the
magnitude in the timescape model has previously been overlooked due to a
prejudice that there can be no significant time dilation for weak fields. Such
physical intuition is driven by our experience of strong field gravity for
isolated systems, in which there is a well-defined canonical clock at spatial
infinity. However, there are no isolated systems in the actual Universe,
and the problem of calibrating the effective asymptotic clocks in regions
of different densities in the absence of a timelike Killing vector is a
qualitatively new one.

In the timescape scenario the relative volume deceleration of regions
of different density is treated as a physical parameter in the normalization
of average asymptotic clocks. Although the relative deceleration is
instantaneously tiny (typically $\lsim10^{-10} $m$\;$s$^{-2}$ \cite{equiv})
and well within the weak field regime, when integrated over the lifetime of the
Universe this can cumulatively lead to present epoch clock rate differences of
order 35\%. In particular,
the present Universe is void--dominated whereas structure formation ensures
that all actual observers, and objects which can be observed, are in regions
of greater than critical density, giving us a mass--biased view of the
Universe. There is a difference of calibration of the canonical clocks of
comoving observers in volume--average environments (in voids) where there is no
detectable matter relative to clocks in bound systems, which can be observed.
Cosmic acceleration turns out to be an apparent effect, due to both: (i) the
backreaction of inhomogeneities in changing the average evolution from a FLRW
model; and (ii) the relation of the time parameter which best describes
statistically average cosmic evolution to the physical clocks of observers
in bound systems \cite{clocks}. The cosmic coincidence problem is readily
solved since apparent cosmic acceleration begins when the void volume
fraction, $\fv$, reaches a particular value $\goesas59$\% \cite{clocks}.

After a variety of observational tests \cite{lnw,sw,grb,dnw} the timescape
model remains a viable alternative to the standard model. On large scales its
expansion history is so close to that of the standard model that differences
in luminosity distances are at the level of current systematic uncertainties
in type Ia supernova data \cite{sw}; which model fits better depends on the
method by which supernova light curves are reduced. On small scales, a study
motivated by the timescape model has led to the discovery that the spherically
averaged expansion of the Universe below the scale of statistical
homogeneity is significantly more uniform in the rest frame of the Local
Group of galaxies than in the standard CMB rest frame \cite{hvar}, a result
which is very difficult to reconcile with the standard model.

To fully compete, observational tests of the timescape scenario need to be
developed to a similar extent to the standard cosmology. One particularly
important test is the detailed fitting of the acoustic peaks in the CMB
anisotropy spectrum. Thus far we have successfully fit the angular diameter
distance of the sound horizon at decoupling, which controls the overall
angular scale of the acoustic peaks, thereby constraining cosmological
parameters \cite{lnw,dnw}. However, a fit of the ratios of the peak heights
has not yet been performed. This paper will remedy that situation: we will
perform the first detailed fits of the acoustic peaks in the timescape
scenario.

The problem is a very nontrivial one, since the timescape model revisits
many of the foundational questions of cosmology from first principles,
and much has to be rebuilt from scratch. The timescape scenario
is based on a particular physical interpretation of the Buchert
averaging formalism \cite{buch00}, and ideally in considering the early
radiation dominated universe we should begin with the version of the
formalism that directly includes the effects of pressure in the averaging
procedure \cite{buch01}.

A from--first--principles investigation of the Buchert formalism in the
radiation--dominated epoch using the constraining principles of the timescape
scenario is a huge challenge, however. Thus we will adopt the same
simplification that was made in our recent paper \cite{dnw}, in which we assume
a standard perturbed FLRW evolution at early epochs which is smoothly matched
to a matter plus radiation solution in which the effects of backreaction from
the radiation fluid are neglected in determining background average cosmic
evolution. Only the density gradients of nonrelativistic matter (both baryonic
and nonbaryonic) are assumed to contribute to cosmic backreaction.

Such an approximation is justified by the fact that the Universe was
definitely close to being homogeneous and isotropic at early times. This
approximation also allows us to make use of standard CMB codes, to the extent
that all quantities must be calibrated to late epoch evolution from the
timescape model, which in itself requires considerable recoding, as will
be described.

The plan of this paper is as follows: in Sec.~\ref{cmb_meth} we briefly
review the key features of the standard CMB acoustic peak analysis in the
FLRW and LTB models. In Sec.~\ref{cmb_ts} we extend this methodology to the
timescape cosmology. In Sec.~\ref{comp} our numerical computation strategy is
discussed. In Sec.~\ref{results} we present our key results of fitting the
timescape cosmology to the CMB acoustic peaks with the Planck satellite data.
Section \ref{dis} contains a concluding discussion.

\section{CMB anisotropy overview}\label{cmb_meth}
\subsection{The standard FLRW model}\label{cmb_std}

The standard FLRW models are phenomenologically highly successful, and
much of this success is built on their application to the problem of the
evolution of perturbations in the early Universe, and the observable
signatures of these perturbations as temperature fluctuations of order
$\DE T/T\goesas10^{-5}$ in the mean CMB temperature, $T=2.7255\,$K. The
tools that have been developed for the analysis of CMB anisotropies, and the
theoretical framework on which they are based, have taken decades of
development. Since we assume that the standard approach is a good
approximation at early times, we will adopt standard tools where possible.
It is therefore useful to firstly recall the key features of the standard
approach, in order to describe which features will remain unchanged and
which will be revisited.

In the standard approach CMB anisotropies are determined from density
fluctuations in the early Universe, which are calculated numerically by
tracing the time evolution of the distributions of baryons, nonbaryonic Cold
Dark Matter (CDM) and photons. The starting point is Boltzmann's equation
\begin{equation}
\hat L[f] = C[f] \label{Bol}
\end{equation}
for the time evolution of each particle distribution function, $f$, where
$\hat L$ is the relativistic Liouville operator on the particle's phase space
and $C$ is a collision term describing particle production and destruction.
The total time rate of change can be evaluated along the particle geodesics of
a FLRW geometry with first order perturbations, such as (\ref{FLN}) in the
case of the Newtonian gauge. The Einstein field equations couple the
distributions of all the interacting constituents into a system of
differential equations generally referred to as the Einstein-Boltzmann
equations. The presence of the collision term means that particle numbers
are not necessarily conserved in phase space.

The principal epochs that are of relevance
can be summarized as follows.

\begin{enumerate}
\item {\em Inflation}. Following an early period of exponentially rapid
expansion and particle production, initial quantum fluctuations manifested
themselves as perturbations in the densities and bulk velocities of matter
fields in the energy momentum tensor. By Einstein's equations these
fluctuations give rise to gravitational
potentials which perturb the background metric.
\item {\em Big Bang Nucleosynthesis (BBN)}. As the Universe expanded and
the temperature dropped, the light elements $^2$H, $^3$He, $^4$He, and
$^7$Li were formed by a series of nuclear reactions, with final abundances
relative to hydrogen which are largely determined by the baryon--to--photon
ratio, $\etBg$, but also by the effective number of neutrino species $\Nn$.
\item {\em Acoustic oscillations}. Before the time of last scattering and
during recombination the Thomson scattering of electrons and CMB photons
facilitated the transfer of energy and momentum between these
species, and also between CMB photons, protons and other charged light nuclei.
The series of peaks
and troughs seen in the CMB power spectrum today arise from the acoustic
oscillations that propagated through the plasma of electrons, protons and CMB
photons in this era. The $13.6\,$eV binding energy of hydrogen restricts the
amplification or diminution of the amplitude of the acoustic waves in the
primordial plasma.
\item {\em Reionization}. After recombination the direct mechanism of
photon--electron scattering which alters CMB photon energies stopped until the
Universe was once again ionized by the bursts of radiation from the formation
of the first stars. The imprint left on the CMB anisotropy spectrum from
reionization is comparatively marginal, featuring principally as a reduction in
amplitude which does not change other salient features.
\item {\em Propagation through intervening structures.}
Our observations of the CMB temperature anisotropies depend on the propagation
of photons over the entire period from last scattering until today. Photon
geodesics are affected both by the background cosmology, and the deviations from
the background due to the growth of structure. One key effect is the
{\em late time integrated Sachs--Wolfe (ISW) effect}: when CMB photons traverse
a region where the gravitational potential changes over time, the boost in
energy when CMB photons fall into a potential well is not canceled by the
reduction in energy when the photons climb out of the potential well. This
has a significant effect for large angles in the anisotropy spectrum.

The power spectrum is also affected by a number of secondary effects
caused by cosmic structures, including the Sunyaev-Zel'dovich effect, weak
gravitational lensing and changes to photon polarization. In this paper,
however, we are simply concerned with the primary anisotropies.
\end{enumerate}

Within a specific FLRW model the features seen in the CMB power spectrum can
be attributed to the relative magnitudes of the various parameters including:
the Hubble constant, $\Hn$, and the density parameters of all nonrelativistic
matter, $\OmMn$; baryons, $\OmBn$; all radiation species, $\OmRn$; photons,
$\Omgn$; scalar curvature, $\Omkn$; and dark energy, $\OmLn$.

A physical understanding of the features of the CMB anisotropies was made
possible by the semi--analytic methods first developed by Hu and Sugiyama
\cite{Hu_1}, and
later on replicated and refined by numerous other authors
\cite{Weinberg_1,Dodelson_1,Mtext}. A concise exposition can be found in
\cite{Julien_1}, the key point relevant to our discussion of CMB
anisotropies in the timescape cosmology being the following.

\begin{itemize}
\item[(i)] The location of the first peak of the CMB temperature power
spectrum in multipole space depends on the angle $\theta=d_s(\tdec)/
\dA(\tdec)$ at which the sound horizon is seen today. The angular
diameter distance to the last scattering surface\footnote{Here $\mathop{\hbox
{sinn}}(x)\equiv\{\sin(x)$, $\Omkn<0$;\ $x$, $\Omkn=0$;\ $\sinh(x)$, $\Omkn>0
\}$.} $$\dsp\dA(\tdec)={a(\tdec)\,c\over a\Z0\Hn|\Omkn|^{1/2}}\mathop{\hbox
{sinn}}\left[|\Omkn|^{1/2}\int_0^{z_{\rm dec}}{\Hn\,\dd z\over H}\right]$$
depends on the expansion rate and can change substantially with changes in
$\OmLn$, $\Omkn$, and $\Hn$. The proper scale of the sound horizon $d_s(t)=a(t)
\int_0^tc_s\dd t/a$ is changed when $\OmBn$, $\OmMn$ and $\OmRn$ are changed.
\item[(ii)] The peaks and troughs in the power spectrum arise from constructive
and destructive interference of the sound waves in the baryon--photon fluid.
This interference is not exactly in phase or out of phase and the resulting
ratios of odd to even peaks depend on $\OmBn/\Omgn$ (or equivalently on
$\etBg$ the baryon-to-photon ratio) that
determine the relative phase of the oscillating waves.
\item[(iii)] On small angular scales, or equivalently for large multipoles,
$\ell\sim(k/a)\dA$, the CMB photons can diffuse and wash away anisotropies.
This is apparent as a decaying of the CMB power spectrum
amplitude as the multipole moments increase.
In the hydrodynamic limit, with the energy momentum tensor taken to be that of
an imperfect fluid, this damping of sound waves for wavelengths smaller than
the diffusion length
\[\qquad\
\lambda_d^2 = a_L^2 \int_0^{t_L}\!\!\frac{1}{6a^2(1+R)\siT n_e}\left\{
\frac{16}{15}+\frac{R^2}{(1+R)} \right\}c\,\dd t
\]
is due to viscosity and heat conduction in the baryon--photon plasma. Here
$R=(3/4)\rho_B /\rho_{\gamma}$, $\siT$ is the Thomson cross section and
$n_e$ is the number density of free electrons. The power spectrum at a given
multipole is damped by a factor
\[
\exp\left(-\frac{k^2}{a^2}\lambda_d^2\right)\approx\exp\left(-\frac{\ell^2}
{\dA^2}\lambda_d^2\right)\,.
\]
The parameters $\OmBn$, $\OmMn$, $\etBg$ modify the power spectrum by changing
$\dA$ and $\lambda_d$ in the damping factor.
\item[(iv)] The influence of reionization is to suppress the power spectrum
by a factor $\exp(-\taur)$, where the optical depth $\tau=\int
\siT n_e\,\dd t$ is evaluated at the
reionization epoch. With the exception of
very small multipoles the rescaling is $\Cl\obs\to\exp(-2\taur)\Cl\obs$.
\item[(v)] The overall amplitude of the power spectrum is proportional to the
amplitude $A_s$ of the primordial perturbations. In this paper we use the
high-$\ell$ Planck power spectrum data $\ell\ge50$ and for these multipoles the
reionization effect is degenerate with the primordial perturbation amplitude
because the product $\exp(-\taur)A_s$ multiplies the power spectrum.
\item[(vi)] The power spectrum is calculated by convolving the temperature
multipoles $\Theta_\ell$ with the primordial spectrum and integrating over all
Fourier modes
\[
\Cl = \frac{1}{2\pi}A_s \int \frac{dk}{k}\left[\Theta_\ell(t_0,k)\right]^2
\left(\frac{k}{k_0}\right)^{n_s-1}\,.
\]
For small values of $\ell$ (large angles) the temperature multipoles are only
related to the monopole at recombination and the integral can be performed
analytically. For spectral index $n_s=1$, $\Cl \sim 1/[\ell(\ell+1)]$.
This scaling is visible as the Sachs-Wolfe plateau. For $n_s<1$, $\Cl\sim
1/[\ell^{n_s}(\ell+1)]$ and there is more power at these multipoles compared
to the case of $n_s=1$. In general the effect of $n_s$ on the CMB power
spectrum is neither a simple shifting of the peaks nor a simple change in
amplitude.
\end{itemize}

\subsection{Exact inhomogeneous models}\label{cmb_in_inhom_models}
In a step up in complexity from the homogeneous isotropic FLRW solutions the
CMB has been also been studied in the spherically symmetric but inhomogeneous
LTB model \cite{L,T,B}. On small scales this solution is an excellent
approximation for the voids \cite{HV1,HV2,Pan11,minivoids} that dominate
the Universe at the present epoch. Applied to gigaparsec scales the solution
becomes a toy model that is physically unlikely, and which violates the
Copernican principle. Nonetheless, being an exact solution of the Einstein
field equations, it is amenable to direct analytic study, and such studies
have included the investigation of the CMB for gigaparsec voids
\cite{BW08,Biswas_1,Clarkson_1,Zibin_1,NS10}.

The LTB solutions have a dust energy--momentum tensor, and can only
apply at epochs in the matter--dominated era. An early time spherically
symmetric radiation plus matter background which evolves to an LTB solution
is beyond the realm of current investigations. Consequently,
the study of CMB anisotropies in LTB cosmologies has to date used models which
initially coincide with a FLRW model. The primary temperature anisotropies
are evolved using the Boltzmann hierarchy with a FLRW background at early times
and the resulting CMB power spectrum is then modified to account for
differences in expansion rate, matter densities and angular diameter distances
between the FLRW and LTB background solutions
\cite{BW08,Biswas_1,Clarkson_1,Zibin_1,NS10}.

Since the timescape model is close to a FLRW model at early times, we will
adopt a similar approach in this paper. Specifically we will adapt the method
outlined in \cite{MRD10}, in which it is shown that the CMB power spectra in
two different models can be mapped onto each other as follows.

Consider two cosmological models with angular diameter distances $\dA'$ and
$\dA$ and a common proper length scale $L$ at the surface of last scattering.
This is viewed at an angle $\theta'= L/\dA'$ and $\theta=L/\dA$ respectively in
the two models. Vonlanthen, R\"as\"anen and Durrer \cite{MRD10} show that the
CMB angular power spectra in these models are related via the integral
\beq
\Cl=\sum_{\tl}\frac{2{\tl}+1}{2}C_{\tl}'\int_0^{\pi}\!
\sin\theta\,\dd\theta\,P_{\tl}\left[\cos(\theta\,\dA/\dA')\right]P_\ell
(\cos\theta). \label{Durrer_approx_1}
\eeq
Eq.\ \eqref{Durrer_approx_1} is derived using the assumption that apart from
the overall amplitude of the two CMB power spectra, the only differences
between the two spectra are due to the differences in the distance to the
surface of last scattering characterized by $\dA'$ and $\dA$. Furthermore,
for high multipoles ($\ell\ge 50$)
Vonlanthen, R\"as\"anen and Durrer \cite{MRD10} show that
(\ref{Durrer_approx_1}) can be approximated as
\beq
\Cl\approx\left(\frac{\dA'}{\dA}\right)^2 C_{\frac{\da'}{\da}\ell}'\;.
\label{Durrer_approx_2}
\eeq
Let us refer to the cosmology with $\dA'$ as a {\em reference model}. Then
the $\Cl$ in the second model at any multipole $\ell$ is found from the scaled
$\tl=\ell\,\dA'/\dA$ of the reference model.
Zibin, Moss and Scott \cite{Zibin_Moss_Scott} first derived
\eqref{Durrer_approx_2} using a different method and the revised results in
\cite{CFZ09} also agree with (\ref{Durrer_approx_2}).

\section{CMB anisotropy overview for the timescape model}\label{cmb_ts}

We will use \eqref{Durrer_approx_2} to calculate the CMB power spectrum in the
timescape model from a reference FLRW power spectrum. In doing so, we must
implicitly assume that the key features of the CMB anisotropy spectrum
are close to those of a FLRW model apart from the shift factor. In particular,
we neglect:
\begin{itemize}
\item[(i)] the effects of backreaction on average cosmic evolution in the
radiation--dominated epoch \cite{buch01};
\item[(ii)] differences in the late time ISW effect between the timescape and
FLRW models.
\end{itemize}
Even if the timescape model is close to a FLRW model in the radiation dominated
epoch, the detailed treatment of the late time ISW effect may differ. However,
this might be expected to only affect large angle multipoles ($\ell\lsim50$),
and we will not include these multipoles in fitting the Planck data.

Our method of determining an appropriate FLRW reference model is
complicated by the fact that in the timescape scenario one is not dealing
with a single set of cosmological parameters common to all observers. In the
presence of inhomogeneity there is more than one single class of average
observers, and in general there are both bare and dressed cosmological
parameters \cite{BC02,BC03}. We will first briefly
review the details of the relationship between bare and dressed parameters
in the timescape model, and then describe how the methodology of
Sec.~\ref{cmb_meth} is modified to find reference FLRW models, whose
expansion history from last scattering until the present is closest to
that of the timescape scenario.

\subsection{Cosmological parameters in the timescape scenario}

We follow Ref.~\cite{dnw} in considering a universe containing nonrelativistic
matter plus radiation (photons and neutrinos), of respective densities
$\rhM$ and $\rhR$, whose evolution is governed by the
Buchert equations \cite{buch00,buch01}. The radiation pressure $\pR
=\frac13\rhR c^2$ is assumed to commute under the Buchert
average\footnote{Here $\vt$ is the expansion scalar and angled brackets denote
the spatial volume average of a scalar quantity on the surface of average
homogeneity, so that $\ave{\pR}\equiv\left(\int_\DD\dd^3x\sqrt{\det\gd}\,{\pR}
(t,\bx)\right)/\Vav(t)$, where $\Vav(t)\equiv\int_\DD\dd^3x\sqrt{\det\gd}$ is
the average spatial volume, $\gd_{ij}$ being the 3-metric. The domain $\DD$ is
taken to be the particle horizon volume in our case.} \cite{buch00},
\beq
\pt_t\ave{\pR}-\ave{\pt_t \pR}=\ave{\pR\vt}-\ave{\pR}\ave{\vt}=0,
\eeq
throughout the evolution of the Universe, so that it is solely the
growth of gradients in the nonrelativistic matter density which drives
the growth of inhomogeneity in the Universe and backreaction on
average cosmic evolution as compared to a FLRW model. The present epoch
horizon volume, $\Vav=\Vav\ns i\ab^3$, is assumed to be a statistical
ensemble of disjoint void and wall regions characterized by respective scale
factors $\av$ and $\aw\,$, which are related to the volume--average scale
factor by
\beq
\ab^3=\fvi\av^3+\fwi\aw^3.\label{avs}
\eeq
Here $\fvi$ and $\fwi = 1-\fvi$ represent the fraction of the initial volume,
$\Vav\ns i$, in void and wall regions respectively at an early unspecified
epoch. If one defines $\fw(t)=\fwi \aw^3/\ab^3$ to be the {\em wall volume
fraction}, and $\fv(t)=\fvi\av^3/\ab^3$ the {\em void volume fraction}, then
(\ref{avs}) may be rewritten as
\beq
\fv(t)+\fw(t) = 1.
\eeq
The voids are assumed to have negative spatial curvature characterized
by $\Rav\ns{ v}\equiv 6\kv c^2/\av^2$ with $\kv<0$, while wall regions
\cite{clocks} are on average spatially flat, $\Rav\ns{ w}=0$.

With these assumptions the independent Buchert equations may be written as
\cite{dnw}
\bea
&\OMM+\OMR+\OMk+\OMQ=1,\label{beq1}\\
&\dsp\ddot\fv+{\dot\fv^2(2\fv-1)\over 2\fv(1-\fv)} +3\dot\fv\bH-\frac{3}{2}
(1-\fv)\OMk\bH^2=0,\label{beq2}
\eea
where
\beq
\bH\equiv{\dot\ab\over\ab}=\fw\Hw+\fv\Hv\,,\label{bareH}
\eeq
with $\Hw\equiv\dot a\ns{w}/\aw$ and $\Hv\equiv\dot a\ns{v}/\av$. We describe
$\bH$ as the
{\em bare} or {\em volume--average} Hubble parameter and
\bea\OMM&=&{8\pi G\rhb\Z{M0}\abn^3\over 3\bH^2\ab^3}\,,\label{om1}\\
\OMR&=&{8\pi G\rhb\Z{R0}\abn^4\over 3\bH^2\ab^4}\,,\label{om2}\\
\OMk&=&{\al^2\fv^{1/3}\over \ab^2\bH^2}\,,\label{om3}\\
\OMQ&=&{-\dot\fv^2\over 9\fv(1-\fv)\bH^2}\,,
\label{om4}
\eea
are the {\em bare} or {\em volume--average} density parameters of
nonrelativistic matter, radiation, average spatial curvature and kinematic
backreaction\footnote{The kinematic backreaction $\QQ\equiv\frn23\left(\langle
\vt^2\rangle-\langle\vt\rangle^2\right)-2\langle\si^2\rangle$ includes a
contribution from the shear scalar, $\si^2=\frn12\si_{\al\be}\si^{\al\be}$,
in general. We assume this to be negligible in cosmic averages, so that the
backreaction is determined solely by the variance in volume expansion between
walls and voids. One then finds $\QQ=2\dot\fv^2/[3\fv(1-\fv)]$.} respectively.
Here $\al^2\equiv-\kv c^2\fvi^{2/3}>0$.

Apart from the presence of the
backreaction term, $\OMQ$, one key difference from the FLRW model is that the
curvature parameter $\OMk$ does not scale simply in proportion to $(\ab\bH)^
{-2}$; i.e., cosmic expansion does not preserve average spatial curvature. This
leads to important phenomenological differences from the FLRW models, both
in the timescape scenario \cite{clocks}--\cite{dnw} and in other approaches
\cite{rob13}--\cite{rbof14} to backreaction. The nonrigid evolution of the
average spatial curvature is subject to a cosmological test \cite{cbl07}, for
which current observational bounds \cite{smn14} are consistent with the
timescape scenario.

The bare density parameters (\ref{om1})--(\ref{om4}) are fractions of the
critical density, $\bar\rh\ns{cr}=3\bH^2/(8\pi G)$, in terms of the
volume--average Hubble parameter, and the time derivative in
(\ref{beq1})--(\ref{om4}) is the volume--average, or bare, Buchert time
parameter which best describes average cosmic evolution. In the timescape
scenario this time parameter is operationally understood to only coincide with
the clock of an isotropic observer\footnote{An isotropic observer is one who
is understood to see a CMB sky which is as close to isotropic as possible.}
whose local regional density coincides with the
particle horizon volume average density $\ave{\rh}$. On account of a
relative volume deceleration, this time parameter is not generally
synchronous with the clocks of other isotropic observers for whom the
locally isotropic regional density is different.

Observers in bound systems are necessarily in regions which are locally
greater than critical density, and can be always be enclosed in compact
{\em finite infinity} surfaces which are marginally expanding at the boundary
but within which the average expansion is zero, and average spatial curvature
is zero \cite{clocks}. The volume bounded by the union of these surfaces
defines the wall regions discussed above. While the wall regions are always
expanding less quickly than the voids, as measured by any one set of clocks, so
that \beq h_r\equiv\Hw/\Hv<1\,, \eeq
the timescape model embodies the physical principle that in the
statistical geometry one can always make a choice of rulers and clocks
to make the expansion uniform \cite{equiv}.

Clocks of isotropic observers in the denser wall regions, which experience
greater regional volume deceleration, run slower relative to those in less
dense void regions. With this condition the clocks of wall observers are found
to keep {\em wall time}, $\dd\tw=\dd t/\gb$, where
\beq
\gb=1+\left(1-h_r\over h_r\right)\fv
\eeq
is the {\em phenomenological lapse function}. Since $\OMQ=-(1-\fv)(1-\gb)^2/
[\fv\gb^2]$, from (\ref{beq1}) it follows that the phenomenological lapse
function is related to the bare density parameters at any epoch
by\footnote{This corrects a typographical error in Eq.~(22) of \cite{dnw}.}
\beq
\gb={\sqrt{1-\fv}\,\Bigl[\sqrt{1-\fv}+\sqrt{\fv(\OM-1)}\;\Bigr]
\over1-\fv\OM}\,,
\label{gb1}\eeq
where
\beq \OM\equiv1-\OMQ=\OMM+\OMR+\OMk\,.\label{gb2}\eeq

In \cite{clocks} a radial null geodesic matching procedure is implemented
to determine effective dressed cosmological parameters inferred by wall
observers, such as ourselves, who mistakenly assume that our regional spatially
flat geometry extends to the entire Universe. The dressed Hubble parameter
is found to be
\beq H=\gb\bH-\gb^{-1}\dot\gb\ \,,\label{Hd}\eeq
with a present epoch value that corresponds to the Hubble constant determined
on scales larger than the statistical homogeneity scale ($\gsim100\hm$).

One can similarly define a dressed matter density parameter $\Omega\Ns{M}
\equiv\gb^3\OMM$ which takes a numerical value closer to the corresponding
parameter for the concordance \LCDM\ model when evaluated at the present epoch.
However, it is just a convenient parametrization and cannot be exactly
identified with the FLRW parameter -- in particular, dressed density
parameters do not constitute a set with values that add up to one as in
the case of the bare parameters.

Many details of further cosmological tests and parameter estimates are given in
\cite{obs}.

\subsection{Calibrating CMB anisotropies}

With our ansatz for the statistical ensemble of walls and voids, and initial
conditions consistent with the amplitude of density perturbations inferred
from the CMB, it turns out that at any instant the magnitudes of the bare
density parameters in (\ref{beq1}) do not differ vastly from those of {\em
some} Friedmann model. For example, for typical solutions \cite{dnw} at last
scattering the void fraction is tiny, $\fv\goesas2\times10^{-5}$ and
$\OMQ\goesas-1\times10^{-5}$.

The void fraction grows considerably over time, and backreaction grows in
amplitude as the voids overtake the walls by volume, but it then subsequently
decreases and its amplitude is bounded by $|\OMQ|<0.043$. Over very small
time periods there is a ``closest Friedmann
model'' in volume--average time. However, given a backreaction term with a
magnitude at the few percent level, and a nonrigidly scaling curvature
parameter, the overall time evolution does differ very significantly from any
single FLRW model without dark energy over long time scales.

Since $|\OMQ|\lsim10^{-5}$ up to last scattering, it seems reasonable to expect
that the physics of the early Universe is little changed in determining
the acoustic oscillations in the plasma, or any earlier processes. In
determining a reference FLRW model for applying (\ref{Durrer_approx_2})
there are two important considerations:
\begin{itemize}
\item Since the equations (\ref{beq1}), (\ref{beq2}) which are closest
to the Friedmann equations are statistical equations in volume--average
time, the calibration of the relevant parameters in the early Universe
in relation the observations of observers who measure wall time
(such as ourselves) has to be carefully considered.
\item To apply any existing CMB anisotropy code a reference FLRW model
has to be found, whose expansion history evaluated at the present day
relative to last scattering is the closest to the expansion history of
a solution to (\ref{beq1}), (\ref{beq2}) as determined by a volume--average
observer.
\end{itemize}

Let us first consider each of the epochs listed in Sec.~\ref{cmb_std} in
relation to the question of calibration of parameters.
\begin{enumerate}
\item {\em Inflation}. The timescape model assumes the phenomenology of
inflationary models and their predictions for the spectrum of density
perturbations up to last scattering. The fact that the Universe does not
evolve by the Friedmann equation after last scattering means that the usual
tight bounds which are often assumed to apply to the root mean square density
contrast $\de\rh/\rh(t)$ on scales at late epochs no longer apply, and this
quantity can reach values of $6$--$8$\% on arbitrarily large scales by the
present epoch \cite{equiv,dust,bscg}. However, this does not directly affect
measurements of the CMB anisotropy spectrum.
\item {\em Big bang nucleosynthesis}. There are no changes to BBN physics.
However, a key parameter in determining BBN rates and subsequent light
element abundances is the baryon--to--photon ratio, $\etBg$. In the timescape
scenario, on account of the large gradients in spatial curvature between
voids and walls and the different relative clocks to which the frequency of
a photon is compared, a volume--average isotropic observer will infer mean
CMB temperature, $\Tb$, which differs from that of a wall observer according
to
\beq
\Tb=\gb^{-1}T \,,
\eeq
at any epoch. In particular, at the present epoch volume--average
temperature $\Tbn=\gbn^{-1} 2.7255\,$K will be up to about 35\% cooler.
This recalibration of volume--average parameters changes the constraints
resulting from light element abundances. In particular, for observational
tests performed to date \cite{lnw,dnw} it was possible to find a best fit
with a volume--average baryon--to--photon ratio, $\etBg$, which avoids a
primordial lithium abundance anomaly \cite{bbn,pill}.

\item {\em Acoustic oscillations}. There are no changes to the physics
governing the acoustic oscillations since the effects of backreaction are
neglected in the early Universe. However, since $\etBg$ may be recalibrated
cosmological parameters relevant to the determination of spectral features
-- in particular, the ratio of nonbaryonic CDM to baryonic matter, $\OmCn=
(\OmMn-\OmBn)/\OmBn$ -- may differ from FLRW--model values. As the ${}^4$He
abundance is potentially changed, this must also be accounted for in the
recombination code.
\item {\em Reionization.} In this paper we do not assume any differences
relative to a perturbative FLRW model calibrated by the timescape parameters.
\item {\em Propagation through intervening structures.} The dressed timescape
geometry \cite{clocks,obs,dnw} is used to determine the angular diameter
distance of the sound horizon, and any other average observational quantities.
In addition to changing the average expansion history, it is very possible
that quantities such as the amplitude of the late time ISW effect may also
differ between the timescape cosmology and the standard \LCDM\ cosmology.
However, this will be neglected here; only small angle multipoles $\ell>50$
will be used in fitting the Planck data.
\end{enumerate}

\subsection{Matching volume--average expansion history to a FLRW model}
\label{mEH}

Since the expansion history in volume--average time is assumed to be close
to that of a FLRW cosmology in the early Universe, and since it is physical
processes up to the decoupling and baryon drag epochs which are principally
responsible for the observed acoustic peaks, we will aim to adapt standard
numerical codes as far as is possible. However, even though the relevant
physical processes occur in the early Universe, standard codes for determining
the acoustic peaks are calibrated to cosmological parameters, such as $\Hn$,
$\OmMn$, $\OmLn$ etc which are evaluated at the present epoch. These are
related to the physical parameters in the early Universe assuming the
expansion history of a FLRW model. Most significantly, the perturbation
equations used as the basis of the Boltzmann hierarchy use a single global
background FLRW geometry, which is written in terms of present epoch
cosmological parameters.

Replacing the expansion history of the FLRW model by that of the timescape
model would require rewriting almost all codes from scratch, a mammoth task
which is not easily realizable considering the decades of effort that have
resulted in codes such as CAMB \cite{camb} and CLASS \cite{class_I,class_II}.
We therefore adopt the approach of determining
the acoustic peaks from a FLRW model with scale factor $\hat a$, whose
expansion history most closely matches that of the volume--average statistical
geometry. This approach is justified since it is the volume--average geometry
whose expansion history describes average cosmic evolution in the sense
closest to the FLRW models. Using the exact timescape solution \cite{dnw}
we can estimate the difference in cosmological parameters at the epoch of
decoupling.

In all cases we will match solutions by determining a FLRW solution for
which \beq{\hat a\Z0\over\hat a}={\abn\over\ab}=1+\bar z\eeq
at all epochs, but with a Hubble parameter which differs in general,
$\hat H\ne\bH$, meaning that the timescape matter and radiation density
parameters (\ref{om1}) and (\ref{om2}) will in general differ from those of
the matched FLRW model. However, we will always arrange the matching so that
$\hat H\simeq\bH$ at early times. At late times we have the freedom to
choose some parameters of the matched FLRW model to be equal to those
of the timescape model at {\em one} epoch. Our choice will be to make the
present epoch Hubble constant, matter and radiation parameters of the
matched model (hatted variables) all equal to the corresponding bare
parameters in the timescape model:
\bea{\hat H}\Z0&=&\Hb \label{m1} \\ \hOmMn&=&\OMMn\label{m2}\\ \hOmRn&=&\OMRn.
\label{m3}\eea
This ensures that matter--radiation equality occurs at the same (bare) redshift,
$\bar z$, in the timescape model as in the FLRW counterpart. Since the
baryonic matter density parameter scales in proportion to the matter
density parameter, it also follows that
\beq\hOmBn=\OMBn.\label{m2B}\eeq
Furthermore, since the Hubble constants are
matched at the present epoch, we have $\hOmMn\hat
h^2=\OMMn\bar h^2$, $\hOmBn\hat h^2=\OMBn\bar h^2$ and $\hOmRn\hat h^2=\OMRn
\bar h^2$, which are parameter combinations typical in the standard FLRW model.
With these choices the present epoch CMB temperature of the FLRW model also
matches the volume--average value $\hat T\Z0=\bar T\Z0=\gbn^{-1}T\Z0$, which
is related to $\hOmRn$ in the standard fashion,
\beq\hOmRn={32\siB\pi G\over3c^3{{\hat H}\Z0}^2}\left[1+\frac78\left(\frac
4{11}\right)^{4/3}\Neff\right]{{\hat T}\Z0}^4\,,\label{Orad}\eeq
where $\siB$ is the Stefan--Boltzmann constant and $\Neff$ is the effective
number of neutrino species.

If we assume that the matching FLRW model is the most general model possible
with both curvature and cosmological constant parameters satisfying
\beq\hOmM+\hOmR+\hOmk+\hOmL=1,\label{flrw}\eeq
then combining (\ref{flrw}) with (\ref{m1}) and (\ref{m2}) at the present epoch
we find
\beq \hOmLn=1-\hOmkn-\OMMn-\OMRn.\label{mIa}\eeq
This places one constraint on $\hOmkn$ and $\hOmLn$, leaving one further
constraint to be found to completely fix the matched FLRW model.

In general, once matter--radiation equality is fixed to be the same in the
two models then $\hat H\simeq\bH$ at all early times when the matched
FLRW equation (\ref{flrw}) and the first Buchert equation (\ref{beq1}) are
both dominated by the matter and radiation densities, with $\hOmM\simeq\OMM$
and $\hOmR\simeq\OMR$. However, there are necessarily small differences
in these parameters given the differences of the other density parameters
appearing in (\ref{beq1}) and (\ref{flrw}). Different choices of the
final matching constraint are found to give differences of magnitude
$\de\hat\Omega<10^{-4}$ in the matched density parameters at decoupling.
We have investigated the following choices (labeled ``A'' for {\em average}
expansion history matching):

\subsubsection{Model {\pAH}: Hubble parameter matched exactly at
decoupling}\label{sssec:H_matched_at_dec}

We can choose the Hubble parameters of the two models to be {\em exactly}
equal at any one particular early time redshift.
Let us match the Hubble parameters at decoupling. Since we already have the
condition (\ref{m1}) this further constraint ensures that the Universe has
decelerated by the precisely the same amount from decoupling until the present
epoch in the matched FLRW model as in the volume--average geometry
of the timescape model.

Equating $\hat H\ns{dec}=\bH\ns{dec}$ and evaluating (\ref{flrw}) at
decoupling we have
\beq\hdec^{-2}\left(\hOmLn+\hOmkn\bxd^2+\OMMn\bxd^3+
\OMRn\bxd^4\right)=1\;,\label{mIb}\eeq
where $\hdec\equiv\bar H\ns{dec}/\Hb$, and $\bxd=\abn/\ab\ns{dec}=1+\bar z\ns
{dec}=\gbn(1+z\ns{dec})/\gb\ns{dec}$, and $\hOmLn$ is given by (\ref{mIa}).
We necessarily have ${\hat\Omega}\ns{k\,dec}<\OM\ns{k\,dec}$ since
$\OMM+\OMR+\OMk=1-\OMQ>1$ in the timescape case, whereas the FLRW parameters
satisfy (\ref{flrw}). For a typical example, such as the best fit
parameters of Ref.\ \cite{dnw}, $\OM\ns{k\,dec}-\hat\Omega\ns{k\,dec}
\simeq1.2\times10^{-5}$, about 16\% of the (small)
value of $\OM\ns{k\,dec}$.

By (\ref{mIb}) the present curvature parameter of the matched
model is fixed to be
\beq
\hOmkn={\hdec^2+\OMMn(1-\bxd^3)+\OMRn(1-\bxd^4)-1\over\bxd^2-1}\,.
\eeq

\subsubsection{Model {\protect\Ar}: Match of effective bare comoving distance
of particle horizon}

As discussed in Sec.~6.4 of Ref.\ \cite{clocks} the solution, $\rb$, of the
averaged Sachs optical equation
\beq
{\ddot\rb}\ +{\dot\ab\over\ab}\mathop{\dot\rb}\ +\Bigl({\dot\fv^2\over3\fv
(1-\fv)}-{\al^2\fv^{1/3}\over\ab^2}\Bigr)\rb=0
\label{optical}\eeq
provides an estimate of the effective comoving scale as would be measured
by a volume average observer, and consequently of an effective volume--average
angular diameter scale $\ave{\dA}=\ab(t)\,\rb(t)\,\de$ for a fiducial
source which subtends an angle $\de$ in its rest frame. In work to date,
we have not used this quantity since it is not directly measured at late
times. Since it averages over both void and wall regions, as a function of
volume--average conformal time, $\eb$, it lies in the range $\eb\le\rb(\eb)
\le\sinh(\eb)$, where the bounds correspond to the FLRW limits: (i)
$\rb(\eb)=\eb$ when $\fv\equiv0$, $\dot\fv\equiv0$; (ii) $\rb(\eb)=\sinh(\eb)$,
when $\fv=\hbox{const}$, $\dot\fv\equiv0$.

For any given timescape parameters, we integrate (\ref{optical}) from
$t=0$ until the present $t=\tn$ to obtain the effective bare comoving
distance of the particle horizon, $\rb\Z{\cal H}$. FLRW parameter values
values $\hOmkn$, $\hOmLn$, are then chosen to simultaneously satisfy
(\ref{mIa}) and the constraint
\beq
\rb\Z{\cal H}=\sinh\left( \int_0^1{\hOmkn^{1/2}\,\dd y\over\sqrt{\hOmLn y^4+
\hOmkn y^2+\OMMn y+\OMRn}}\right),
\eeq
which is the equivalent solution to (\ref{optical}) in the FLRW limit
with $\fv\equiv1$ and $\dot\fv\equiv0$.

\subsubsection{Model {\protect\Ae}: Match of bare conformal time age of the
Universe}

For any given timescape parameters, we integrate $\eb=\int c\,\dd t/\ab$ from
$t=0$ until the present $t=\tn$, to obtain the bare conformal age of the
Universe, $\eb\Z0$. FLRW parameter values
values $\hOmkn$, $\hOmLn$, are then chosen to simultaneously satisfy
(\ref{mIa}) and the constraint
\beq
\eb\Z0=
\int_0^1{\hOmkn^{1/2}\,\dd y\over\sqrt{\hOmLn y^4+\hOmkn y^2+\OMMn y+\OMRn}}\,.
\eeq

\subsubsection{Model {\protect\At}: Match of the bare age of the Universe}

For any given timescape parameters, we determine the age of the
Universe in volume--average time, $\tn$. FLRW parameter values
values $\hOmkn$, $\hOmLn$ are then chosen to simultaneously satisfy
(\ref{mIa}) and the constraint
\beq
\Hb\tn=\int_0^1{y\,\dd y\over\sqrt
{\hOmLn y^4+\hOmkn y^2+\OMMn y+\OMRn}}.
\eeq

\subsubsection{Model {\protect\Ao}: $\hOmLn=0$}

We set $\hOmLn=0$ in the matched FLRW model, so that $\hOmkn$ is simply given
by (\ref{mIa}).

\begin{figure}[!htb]
\begin{center}
\includegraphics[width=8cm,height=6.5cm]{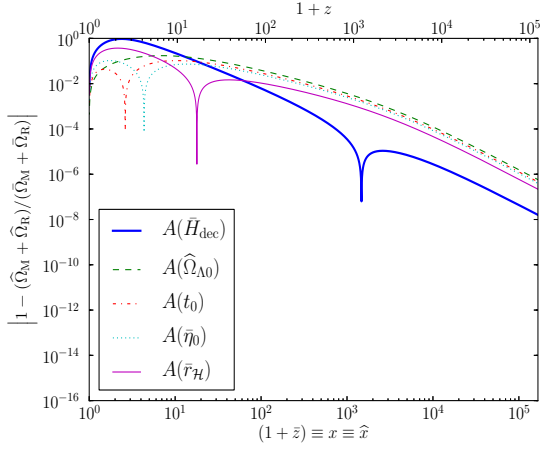}
\end{center}
\caption{\small Fractional difference in the ratio of matter plus radiation
densities of the matched FLRW models relative to the volume--average matter
plus radiation densities of the timescape model, for the average expansion
history matching procedures, using the best fit parameters from
Ref.\ \cite{dnw}.}
\label{fig:densities_diff}
\end{figure}

\subsubsection{Comparison of volume--average expansion history matching methods}

In Fig.~\ref{fig:densities_diff} we show the fractional difference in
the matter plus radiation densities of the matched FLRW models which
satisfy (\ref{m1})--(\ref{m2B}) relative to those of the timescape
model, for the example of the best fit parameters from Ref.\ \cite{dnw},
viz., $\Hn=61.7\kmsMpc$, $\fvn=0.695$, $\etBg=5.1\times10^{-10}$.
The ratio of combined matter and radiation densities
\beq
\frac{\hOmM+\hOmR}{\OMM+\OMR}= \frac{1-\hOmL-\hOmk}{1-\OMQ-\OMk}\label{mr}
\eeq
is initially very close to unity, and in general its departure from
unity is a measure of the extent to which the sum of the dark energy and
spatial curvature parameters in the matched FLRW model differ from the
backreaction and curvature contributions in the timescape model. Model
\AH, which has its final matching condition set at decoupling, has the
smallest difference in the early Universe with $\left|1-\left(\hOmM+\hOmR
\right)/\left(\OMM+\OMR\right)\right|\lsim10^{-5}$ before decoupling. For
the other average expansion history matching methods, which are based on a
matching condition set at late epochs, the corresponding difference is
typically two orders of magnitude larger before decoupling, which still
means a difference of $\lsim10^{-3}$.

\begin{figure}[!htb]
\begin{center}
\includegraphics[width=8cm,height=6.5cm]{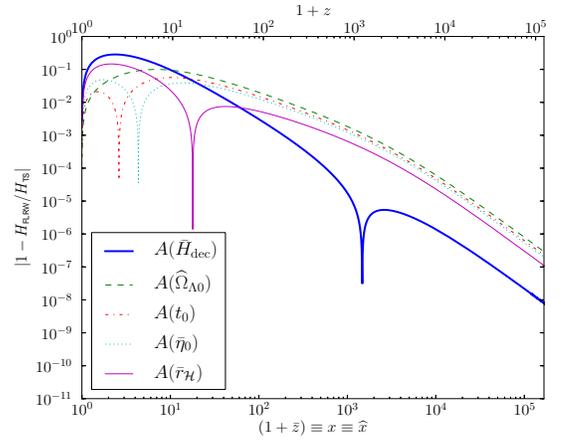}
\end{center}
\caption{\small Fractional difference of the Hubble parameter of the matched
FLRW models relative to the volume--average Hubble parameter of the timescape
model, for the average expansion history matching procedures, using the best
fit parameters from Ref.\ \cite{dnw}. For all procedures ${\hat H}\Z0={\protect
\Hb}$ at the present epoch.}
\label{fig:Hdiff}
\end{figure}

\begin{figure}[!htb]
\begin{center}
\includegraphics[width=8cm,height=6.5cm]{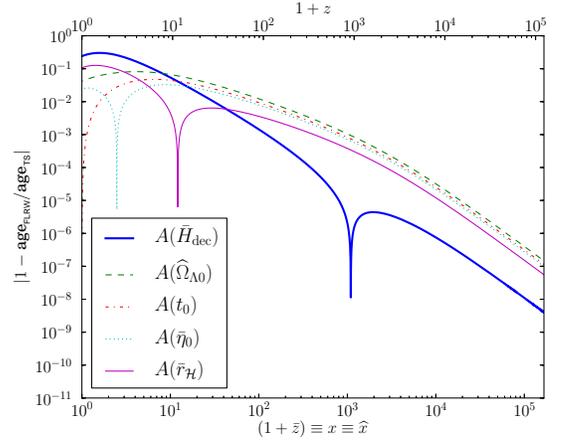}
\end{center}
\caption{\small Fractional difference of the expansion age of the matched FLRW
models relative to the volume--average expansion age of the timescape model, for
the average expansion history matching procedures, using the best fit parameters
from Ref.\ \cite{dnw}. The respective present ages, $\hat t\Z0$, of the matched models
are: {\pAH}, $21.7\,$Gyr; {\protect\Ar}, $19.2\,$Gyr; {\protect\Ae},
$17.9\,$Gyr; \At, $17.5\,$Gyr ($=\tn$); \Ao, $16.8\,$Gyr.}
\label{fig:tdiff}
\end{figure}

\begin{figure}[!htb]
\begin{center}
\includegraphics[width=8cm,height=6.5cm]{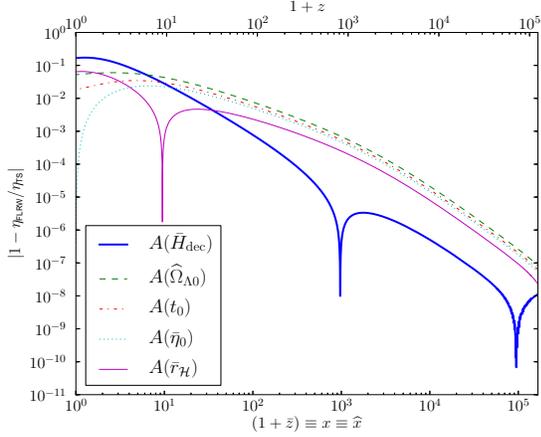}
\end{center}
\caption{\small Fractional difference of the conformal age of the matched FLRW
models relative to the volume--average expansion age of the timescape model, for
the average expansion history matching procedures, using the best fit parameters
from Ref.\ \cite{dnw}. The respective present conformal ages, $\hat\et\Z0$, of
the matched models are: {\pAH}, $4.06$; {\protect\Ar}, $3.70$; {\protect\Ae},
$3.48$ ($={\protect\eb}\Z0)$; \At, $3.41$; \Ao, $3.30$.}
\label{fig:ediff}
\end{figure}

The \AH\ matching method is also seen to produce a closer match at early
times (large redshifts) to quantities such as the volume--average Hubble
parameter, expansion age and conformal time as compared to the other
matching conditions, as is seen in Figs.~\ref{fig:Hdiff}--\ref{fig:ediff}.
This is true for redshifts $\bar z\gsim70$ (bare) or
$z\gsim50$ (dressed). Since the physics of the early Universe is the most
crucial for determining the features of the acoustic peaks, we will therefore
take model \AH\ as the canonical matching method for determining the fit
of the timescape model to the Planck data.

While a general trend of increased departure from FLRW--like behavior can be
seen in all matching types, the largest difference in
Figs.~\ref{fig:densities_diff} and \ref{fig:Hdiff} for the canonical matched
model \AH\ occurs when the backreaction contribution is at its
maximum, $|\OMQ|\simeq0.043$, in the dressed redshift range
$1\lsim z\lsim5$. (See Fig.~1 in \cite{obs} and also Fig.~1 in \cite{dnw}.)
Since the impact of an increase in $|\OMQ|$ on volume--average time and
conformal time is not immediate, but rather cumulative, for the \AH\ matched
model the fractional difference from unity peaks at a lower redshift, as seen
in Figs.~\ref{fig:tdiff} and \ref{fig:ediff}, and then drops again once
$|\OMQ|$ decreases.

\subsection{Matching wall expansion history to a FLRW model}\label{mED}

The matching methods of Sec.~\ref{mEH} ensure that the expansion history
is as close as possible to that of the volume--average expansion history
in one sense or another. It is therefore likely to accurately constrain
the two free parameters that essentially describe the timescape expansion
history, namely the bare Hubble constant, $\Hb$, and the void fraction, $\fvn$,
(or equivalently other pairs of independent parameters such as the dressed
Hubble constant, $\Hn$, and dressed matter density parameter, $\OmMn$).

Although the differences in the density parameters of the timescape model
are small as compared to the matched FLRW at early epochs, (with $\de\hat
\Omega<10^{-4}$), differences which are small in terms of determining the
background may nonetheless be significant in the treatment of the perturbations
of the background, which are a significant part of the standard codes which
we adopt in the early Universe.

The timescape model incorporates the assumption that the geometry on the
finite infinity scale is very close to that of a spatially flat model,
becoming close to Einstein--de Sitter at late times. The relevant perturbation
theory on finite infinity scales is therefore likely to be closer to that of
the spatially flat matter plus radiation solution with no curvature or
$\Lambda$ term.

We will therefore also investigate alternative matching procedures in
which the initial conditions at decoupling are as close as possible to
those of the timescape model, but the expansion history is constrained
to match that of the wall geometry only rather than the volume--average
geometry. We will also consider a second model in which we perform a
similar matching based on the geometry of void centres only. (These
models are labeled ``W'' for {\em wall} expansion history matching.)

\subsubsection{Model {\protect\Ww}: Match of wall expansion history}
\label{mweh}

We construct a spatially flat matter plus radiation model with initial
conditions matched as closely as possible to those of the wall geometry
by taking density parameters
\beq\lOMM+\lOMR=1,\label{EdS}\eeq
with respect to a spatially flat FLRW model with scale factor, $\Ta$, and
Hubble parameter, $\TH$, which are matched so that the physical densities of
the matter, $\Trh\Z M=3\TH^2\lOMM/(8\pi G)$, and radiation, $\Trh\Z R=3\TH^2
\lOMR/(8\pi G)$, are equal to those of the timescape model at decoupling. Thus
\bea\TH\ns{dec}&=&\OMd^{1/2}\,\bH\ns{dec}\label{eqt1}\\
\lOMMd&=&{\OMMd\over\OMd},\label{eqt2}\\
\lOMRd&=&{\OMRd\over\OMd},\label{eqt3}
\eea
where
\beq\OMd=\OMMd+\OMRd=1-\OMQd-\OMkd\,.\label{eqt4}\eeq
On account of (\ref{eqt1}) the expansion of the matched model is very close
to the timescape model as decoupling, but slower by an amount $(\bH\ns{dec}-\TH
\ns{dec})/\bH\ns{dec}<10^{-4}$.

The expansion of the matched model from matter--radiation equality
until decoupling is guaranteed to match that of the timescape model since
\beq {\Ta\ns{eq}\over\Ta\ns{dec}}={\lOMRd\over\lOMMd}={\OMRd\over\OMMd}\,.
\eeq

The spatially flat FLRW model with matter and radiation has a Hubble
parameter given by
\beq
\TH={\TH\ns{eq}\over\sqrt2}\left(\Ta\ns{eq}\over\Ta\right)^2
\sqrt{1+{\Ta\over\Ta\ns{eq}}}\,,\label{th1}
\eeq
while the solution in terms of conformal time $\Te=c\int\dd\tilde t/\Ta$ is
given by \cite{Mtext}
\beq
{\Ta\over\Ta\ns{eq}}=2\left(\Te\over\Te_*\right)+\left(\Te\over\Te_*\right)^2,
\eeq
where $\Te^{-1}_*=\Ta\ns{eq}\TH\ns{eq}/(2\sqrt2\,c)$.
Using (\ref{eqt1})--(\ref{th1}) we find that
\beq
\TH=\bH\ns{dec}\left(\Ta\ns{dec}\over\Ta\right)^{3/2}\sqrt{\OMMd+\OMRd
{\Ta\ns{dec}\over\Ta}}\,.\label{th2}
\eeq
In a similar fashion to the matching methods of Sec.\ \ref{mEH} one final
condition is required for the model matching. In this case we take the
present epoch matched spatially flat FLRW model to match that of the
wall geometry in volume average time \cite{clocks,sol}
\beq
\TH\Z0=H\ns{w0}=\gbn^{-1}\Hb,\label{Hw}
\eeq
given that the wall geometry is extremely close to an Einstein--de Sitter
geometry in volume--average time at late epochs \cite{sol}. As a consistency
check, we find numerically that the present epoch expansion age of the matched
spatially flat FLRW model matches that of the volume--average age of the
Universe in the timescape model.

Combining (\ref{th2}), (\ref{Hw}) we find
\beq
\sqrt{\OMMd\Txd+\OMRd}={\Hb\Txd^2\over\gbn\bH\ns{dec}}\,,\label{th3}
\eeq
where $\Txd\equiv\Ta\Z0/\Ta\ns{dec}$. The expansion of the
matched model until the present epoch is now completely fixed.

The notional present epoch CMB temperature of matched FLRW model is also
fixed as $\widetilde T\Z0=\Txd^{-1}\bar T\ns{dec}$ in terms of the physical
temperature $\bar T\ns{dec}$ of the primordial plasma at decoupling. In
the actual Universe we see CMB photon geodesics which traverse both wall
and void regions. Since the void regions expand faster the expansion from
decoupling until today is always larger in the actual Universe than in the
matched model here, which only has wall regions. Consequently the observed
CMB temperature is always less than the notional $\widetilde T\Z0$.

For computational convenience we note that given a solution $\Txd$ of
(\ref{th3}) the present epoch density parameters of the matched model
are
\bea
\lOMMn&=&{\OMMd\Txd\over\OMMd\Txd+\OMRd},\label{eqt5}\\
\lOMRn&=&{\OMRd\over\OMMd\Txd+\OMRd},\label{eqt6}
\eea
while $\tOM\Z{B0}/\lOMMn=\OMBn/\OMMn$.

\subsubsection{Model {\protect\Wv}: Match of wall expansion history with
initial curvature}\label{sssec:wall_H_matched_at_dec}

Finally, another variation of the matching procedure of Sec.\ \ref{mweh}
is to replace (\ref{EdS})--(\ref{eqt4}) by
\beq\lOMM+\lOMR+\lOMk=0,\label{frw}\eeq
where
\bea\TH\ns{dec}&=&\OMd^{1/2}\,\bH\ns{dec}\label{evt1}\\
\lOMMd&=&{\OMMd\over\OMd},\label{evt2}\\
\lOMRd&=&{\OMRd\over\OMd},\label{evt3}\\
\lOMkd&=&{\OMkd\over\OMd},\label{evt4}
\eea
with
\beq\OMd=\OMMd+\OMRd+\OMkd=1-\OMQd\,.\label{evt5}\eeq
Once again, the expansion between matter--radiation equality and
decoupling, $\Ta\ns{dec}/\Ta\ns{eq}$, matches that of the timescape
model, but the matched model now has a spatial curvature term.

We are effectively still largely matching the expansion history of
the wall geometry, but incorporating an initial negative curvature
consistent with the initial conditions, to see whether this leads to any
noticeable differences.

One may use the Friedmann equation for the matched FLRW model to determine
each of the density parameters $\lOMMn$, $\lOMRn$ and $\lOMkn$ at the present
epoch in terms of their values at decoupling, $\lOMMd$, $\lOMRd$, $\lOMkd$,
and the redshift factor $\Txd=\atn/\at\ns{dec}\equiv1+\widetilde z$. If one
combines the resulting expression with (\ref{evt1})--(\ref{evt5})
one finds
\bea
\lOMMn&=&{\OMMd\Txd\over\Delta},\label{evt6}\\
\lOMRn&=&{\OMRd\over\Delta},\label{evt7}\\
\lOMkn&=&{\OMkd\Txd^2\over\Delta}\label{evt8},
\eea
where \beq\Delta\equiv\OMkd\Txd^2+\OMMd\Txd+\OMRd.\label{evt9}\eeq

Since there is now a constant negative spatial curvature, which is a feature of
neither walls nor voids in the timescape model, there is no appropriate
expansion rate to match to at the present epoch. Instead we will take the
present epoch expansion age of the matched solution to be the age of the
Universe in volume--average time, similarly to the case of Sec.\ \ref{mweh}.
This leads to the constraint
\beq \bH\ns{dec}\tn=\int_0^{\Txd} {u\,\dd u\over\sqrt{\OMkd
u^2+\OMMd u+\OMRd}}\,,
\eeq
which can be used to solve for $\Txd$ and consequently fully constrain
the matched model, with density parameters given by (\ref{evt6})--(\ref{evt9})
and present epoch Hubble parameter
\beq \TH\Z0=\bH\ns{dec}\sqrt{\OMkd\Txd^{-2}+\OMMd\Txd^{-3}+\OMRd\Txd^{-4}}.
\eeq

This concludes the possible matching procedures we have investigated.
There is no equivalent procedure based on the void expansion rate only.
The voids initially form a tiny fraction of the volume and are not
representative of average conditions. At late epochs the voids dominate by
volume, with an expansion law close to that of an empty Milne universe.
An empty universe model cannot be used as the basis of perturbation theory,
however.

\section{CMB analysis: Computational methodology}\label{comp}

Insofar as the details of the evolution of the primordial plasma in
the Universe can be assumed to be given by a standard FLRW model then
for multipoles $\ell>50$ we might expect the matching procedures of
Secs.~\ref{mEH}, \ref{mED} to give reasonably accurate quantitative estimates
of CMB constraints on timescape model parameters.
We therefore use the timescape theoretical CMB temperature power
spectrum which we obtain from each matched FLRW model and
Eq.~\eqref{Durrer_approx_2} with the Planck CMB data \cite{Planck_1} to
constrain timescape parameters.

To find timescape parameters that fit the Planck CMB temperature power spectrum
data \cite{Planck_1} we employ Bayesian analysis to obtain parameter constraints
with the affine invariant\footnote{Unlike the traditional Metropolis--Hastings
\cite{Metropolis} algorithm the affine invariant MCMC algorithm does not require
a knowledge of the covariance matrix of the MCMC parameters and has the
additional advantage of a high acceptance rate. The algorithm only requires the
user to select the initial MCMC values close to the best fit parameters and
provide an estimate of the errors in the parameters. The algorithm then
explores the full parameter space. In a comparison of the Metropolis--Hastings
\cite{Lewis_&_Bridle}, nested sampling and the affine invariant sampling
techniques of Allison and Dunkley \cite{Dunkley_1}, the affine invariant
algorithm is shown to perform very well.} Markov chain Monte Carlo (MCMC)
algorithm \cite{Goodman} using its python implementation \cite{Foreman}.
We fit the parameters of each matched FLRW universe by computing its CMB power
spectrum with the CLASS Boltzmann code \cite{class_I,class_II}. We modified
the code to allow for the different parameter ranges with extended bounds, as
required by the matched FLRW models.

All FLRW model matching procedures also require that we solve the timescape
equations with matter and radiation to find the exact present epoch
timescape parameter values, as described in \cite{dnw}. We must further
extend our previous numerical computations \cite{dnw}, as the baryon--to--photon
ratio, $\etBg$, will now also be included as a base parameters for the MCMC
analysis. In this section we outline the details of all our
numerical computations.

\subsection{Big bang nucleosynthesis}
\nobreak
To date all studies of the timescape model have assumed a range of
values for the baryon--to--photon ratio, $\etBg$, consistent with observed
light element abundances that avoid the lithium--7 abundance anomaly
\cite{bbn,pill}. However, $\etBg$ was not constrained from CMB data. Here we
will constrain $\etBg$ directly from the Planck data. We take $\etBg$ as an
MCMC parameter and treat $h^2\OMBn$ as a derived parameter, in contrast to
\LCDM\ model studies which commonly use $h^2\OmBn$ as a base MCMC parameter.

Since FLRW model calibrations are built into the BBN characterization of
many standard CMB codes, we perform our own BBN calculations by adapting the
{\em fastbbn} code\cite{fbbn0}
of Sarkar and co-workers \cite{fbbn1,fbbn2}. The code takes $\etBg$ and
$\Neff$ as free parameters and determines light element abundances, including
the helium fraction, $Y_{\rm p}$ \cite{fbbn1,fbbn2}. This value of $Y_{\rm p}$
is then used in the timescape recombination calculation, and is also passed to
the Boltzmann code of the matched FLRW model.

Another alternative is to ignore the BBN determination of $Y_{\rm p}$ from
$\Neff$ and $\etBg$ altogether and add $Y_{\rm p}$ to the list of MCMC
parameters. This way $Y_{\rm p}$ is allowed to vary freely and the Planck CMB
data alone set its value. This can be done with $\Neff$ fixed or
free to vary. We have not investigated this last possibility, which if carried
out can be regarded as a consistency check of theoretical BBN predictions.

\subsection{Decoupling and recombination}

To determine timescape parameters we numerically integrate the timescape
equations with matter and radiation, using methods given in Ref. \cite{dnw}.
We also
determine precise values for the redshift of decoupling -- both bare
$\bar z\ns{dec}$ and dressed $z\ns{dec}$ -- and the dressed angular
diameter distance to the last scattering surface at decoupling \cite{clocks}
\beq d\Ns{A\,dec}=\ab\ns{dec}\gb\ns{dec}(1-\fv\ns{dec})^{1/3}\int_{t\ns{dec}}
^{t_0}{\dd t\over\gb\fvf^{1/3}\ab},\label{dAdec}
\eeq
which is needed to shift the CMB power spectrum in multipole space and
also for matching procedures \AH, \Ww\ and \Wv.

In our previous work \cite{dnw} the effects of helium recombination were not
included, and we applied the Saha equation beginning at the end of helium
recombination. To produce more accurate results, we wrote a simple
recombination code that solves for hydrogen and helium ionization fractions,
following methods given in refs.~\cite{Weinberg_1,Chung-Pei,Peebles}, in
terms of the volume--average temperature $\Tb$ in the timescape model. Our
recombination code produces results that are consistent with HyRec \cite{HyRec}
and RECFAST \cite{RECFAST_1,RECFAST_2} when used with a FLRW solution. We
code for helium recombination \cite{Chung-Pei,Peebles} beginning at bare
redshift $\zb=10000$, with the initial helium fraction determined directly
from our own BBN code as the parameters $\etBg$ and $\Neff$ are varied.

\begin{figure}[!htb]
\begin{center}
\includegraphics[width=8cm]{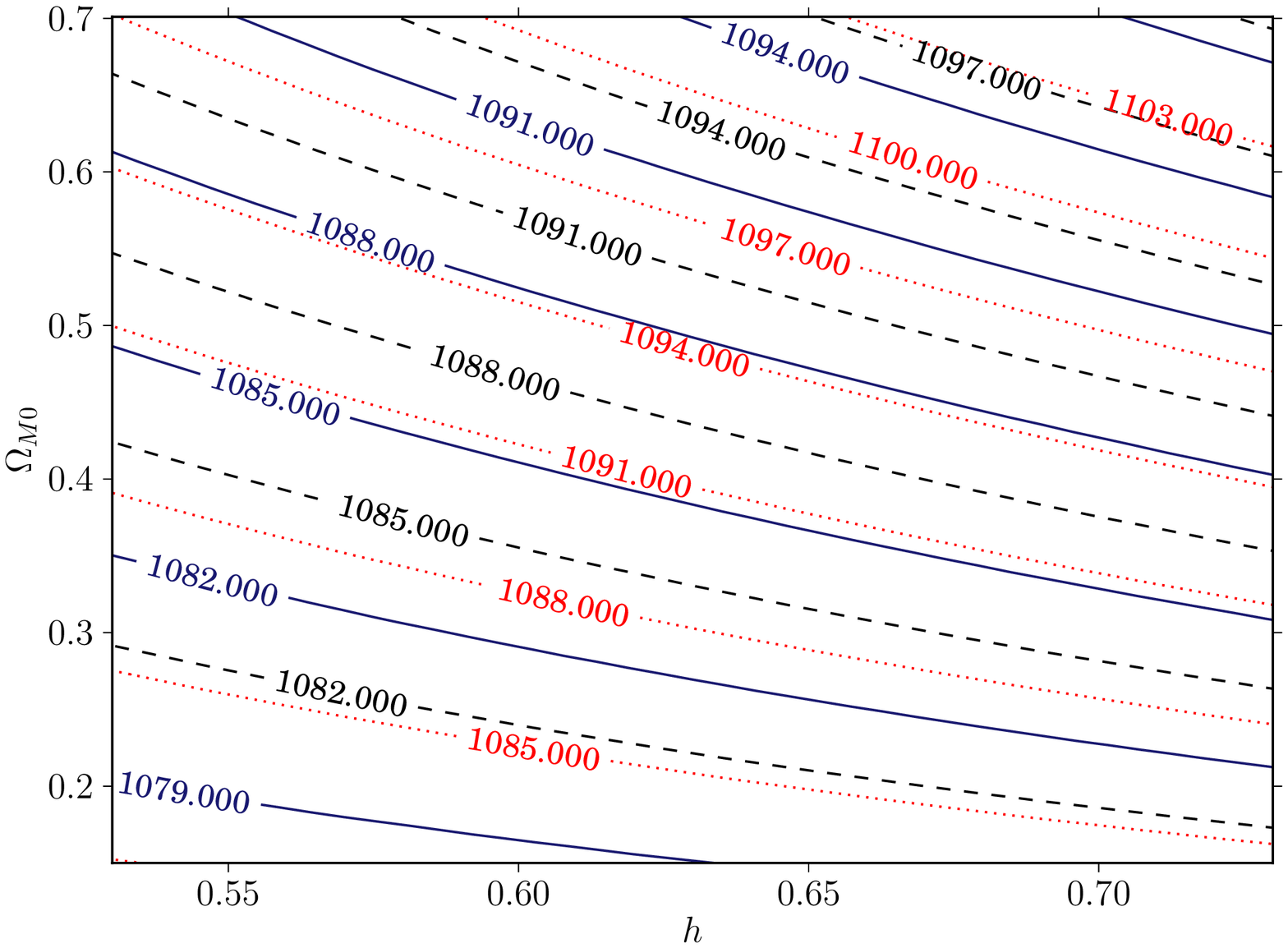}
\end{center}
\caption{\small Contours of dressed redshift of decoupling, $z\ns{dec}$, in
the space of dressed parameters ($h$, $\OmMn$), (where $\Hn=100\,h\kmsMpc$).
Contours are shown for the cases $10^{10}\etBg=$ 5.1 (dotted lines),
6.04 (solid lines) and 6.465
(dashed lines).}
\label{fig:zdec}
\end{figure}
\begin{figure}[!htb]
\begin{center}
\includegraphics[width=8cm]{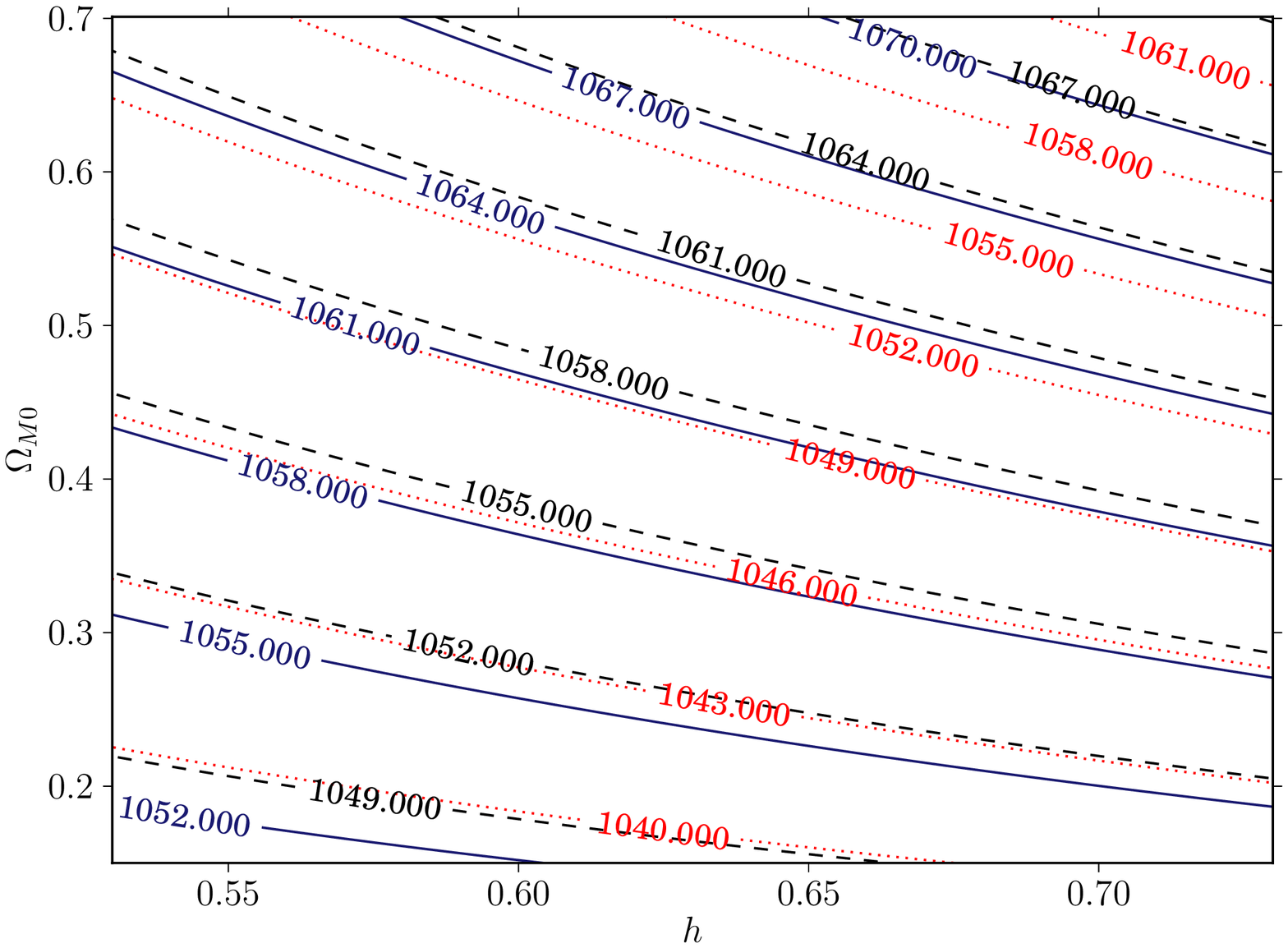}
\end{center}
\caption{\small Contours of the dressed redshift of the baryon drag epoch,
$z\ns{drag}$, in the space of dressed parameters ($h$, $\OmMn$), (where $\Hn=
100\,h\kmsMpc$). Contours are shown for the cases $10^{10}\etBg=$ 5.1
(dotted lines), 6.04 (solid lines) and 6.465 (dashed lines).}
\label{fig:zdrag}
\end{figure}

Including the effects of helium recombination changes the dressed redshift of
decoupling slightly as compared to Ref.\ \cite{dnw}, but has a somewhat
more substantial effect on the estimate of the baryon drag epoch.
This can be seen in Figs.~\ref{fig:zdec}, \ref{fig:zdrag},
where we plot contours of $z_{\rm dec}$ and $z_{\rm drag}$ for three values of
the baryon--to--photon ratio: $\etBg=5.1\times10^{-10}$ as used in Figs.~3 and
4 of Ref.~\cite{dnw}, $\etBg=6.04\times10^{-10}$ which is a best
fit to the \LCDM\ model with the Planck data \cite{Pparm}, and
$\etBg=6.465\times10^{-10}$ which is a best fit for fixed $\Neff=3.046$ using
the canonical \AH\ matching method.

\subsection{MCMC parameters and priors}

In the top part of Table \ref{tb:TS_params_def} we identify the base
parameters, i.e., the timescape parameters varied in the MCMC analysis. All
other timescape parameters are derived from these. Of the 7 base parameters
the 5 parameters $\{\fvn$, $h$, $T_0=2.7255\,$K, $\etBg$, $\Neff\}$ comprise
the set that fixes the background timescape model, while the additional two
parameters $\{n\ns{s}$, $A\ns{s}\}$ are needed for the matched perturbed FLRW
model. The CMB temperature $T_0=2.7255\,$K measured by a wall observer is
constrained by observation \cite{Fixsen}, and is not varied.
%-------------------------------------------------%
\begin{table*}[htb]%p
\footnotesize
\centering
\caption{\label{tb:TS_params_def}\small
The timescape parameters varied in MCMC analysis with their assumed priors are
shown in the top section. The derived parameters relevant to a volume--average
observer are in the middle section and the dressed derived parameters measured
by a wall/galaxy observer are shown in the third group. A brief description of
the parameters is also given.}
\makebox[\textwidth]{
\begin{tabular}{ @{}l c l }
\noalign{\vskip 3pt \hrule height 1.pt \vskip 3pt }
Parameter & Prior range & \hspace{1 in} Description \\
\noalign{\vskip 3pt\hrule height 0.2pt\vskip 3pt}
$\fvn$\dotfill & $[0.3,0.9]$ & Fraction of horizon volume in voids today \\[2.5pt]
$\Hn$ \dotfill & $[0.3,0.9]$ & Dressed Hubble parameter\\[2.5pt]
$10^{10}\etBg$ \dotfill & $[4.0,7.0]$ & $10^{10}\times$ Bare baryon to photon ratio \\[2.5pt]
$\Neff$\dotfill & $[0., 0.6]$ & The effective number of neutrino species \\[2.5pt]
$n\ns{s}$\dotfill & $[0.9,1.1]$ & Scalar spectrum power-law index, pivot $k_0=0.05\,{\rm Mpc}^{-1}$\\[2.5pt]
$10^9A_{\rm s}$\dotfill & $[2.0,20.]$ & Amplitude of the primordial curvature perturbations $(k_0=0.05~{\rm Mpc^{-1}})$\\[2.5pt]
$T_0$\dotfill && CMB temperature of $2.725$K measured by a wall observer \\[2.5pt]
$\left(\OMM/\OMR\right)\ns{dec}$\dotfill & $> 1.$& The ratio of matter to radiation energy density at decoupling \\
\noalign{\vskip 3pt\hrule height 0.2pt\vskip 3pt}
$\Hb$\dotfill && Bare Hubble constant \\[2.5pt]
$\Tb_0$\dotfill && CMB temperature seen by a volume--average observer\\[2.5pt]
$t_0$\dotfill && Age of the Universe (volume--average observer in Gyr)\\[2.5pt]
$\gbn$\dotfill && Present phenomenological lapse function \\[2.5pt]
$Y_{\rm p}$\dotfill && Helium fraction\\[2.5pt]
$\OMBn$\dotfill & & Bare baryon density parameter\\[2.5pt]
$\OMCn$\dotfill && Bare cold dark matter density parameter \\[2.5pt]
$\OMMn$\dotfill && Bare total matter density\\[2.5pt]
$\OMkn$\dotfill && Bare curvature parameter \\[2.5pt]
$\OM_{{\cal Q}0}$\dotfill && Bare backreaction parameter\\[2.5pt]
$\bar z\ns{dec}$\dotfill && Bare redshift of decoupling\\[2.5pt]
$\bar z\ns{drag}$\dotfill && Bare redshift of drag epoch\\[2.5pt]
$\bD_{\rm s}(\bar z\ns{dec})$\dotfill && proper size of sound horizon at $\bar{z}=\bar z\ns{dec}$ (Mpc)\\[2.5pt]
$\bD_{\rm s}(\bar z\ns{drag})$\dotfill && proper size of sound horizon at $\bar{z}=\bar z\ns{drag}$ (Mpc)\\
\noalign{\vskip 3pt\hrule height 0.2pt\vskip 3pt}
$\tau_{\rm w0}$\dotfill && Age of the Universe (galaxy/ wall observer in Gyr)\\[2.5pt]
$\OmBn$\dotfill & & Dressed baryon density parameter \\[2.5pt]
$\OmCn$\dotfill && Dressed cold dark matter density parameter\\[2.5pt]
$\OmMn$\dotfill && Dressed total matter density parameter\\[2.5pt]
$z\ns{dec}$\dotfill && Dressed redshift of decoupling\\[2.5pt]
$z\ns{drag}$\dotfill && Dressed redshift of drag epoch\\[2.5pt]
$100\theta_{\rm dec}$\dotfill && $100\times$ angular scale of sound horizon at $z=\bar z\ns{dec}$\\[2.5pt]
$100\theta_{\rm drag}$\dotfill && $100\times$ angular scale of sound horizon at $z=\bar z\ns{drag}$ \\[2.5pt]
$d_{\rm A}$\dotfill && Dressed angular diameter distance to sound horizon at decoupling (Mpc)\\[2.5pt]
$d_{\rm A,drag}$\dotfill && Dressed angular diameter distance to sound horizon at drag epoch (Mpc)\\
\noalign{\vskip 3pt \hrule height 1.pt \vskip 3pt}
\end{tabular}}
\end{table*}
%-------------------------------------------------%

We use $\Neff=3.046$ for the base timescape model, as our main aim is to
explore the parameter space in the timescape cosmology while remaining
consistent with known particle physics\footnote{Independent of
cosmological constraints on $\Neff$ a recent compilation \cite{Beringer_etal}
of particle physics experiments gives the number of light neutrinos as
$2.984\pm0.008$ for experiments that track $Z$ boson production from $e^+e^-$
annihilation, and $2.92\pm0.05$ for experiments that study the
$e^+e^-\rightarrow\nu\bar\nu\gamma$ process. In cosmology, the convention is
to parametrize the total radiation energy density parameter by (\ref{Orad}).
Instead of $\Neff=3$ the value $\Neff=3.046$ is used \cite{Mangano_etal}
in (\ref{Orad}) to account for noninstantaneous decoupling between
electrons/positions and neutrinos, QED corrections to the photon,
electron/positron plasma and neutrino flavour oscillations assuming three
species of neutrinos.}. The case in which $\Neff$ is left free to
vary is discussed in the Appendix.

For other base parameters we choose a wide prior range to explore the full
parameter space. Flat priors are chosen, with the exception that the ratio
of matter to radiation energy density at decoupling is constrained to be
larger than unity, $\OM\Z{M\,\srm{dec}}/\OM\Z{R\,\srm{dec}}
>1$, as a strict prior. The range of priors for $\fvn$, $h$ are chosen to be
larger than their bounds found in previous studies \cite{sw,dnw}.

We assume that the primordial scalar perturbations are adiabatic with a spectrum
\begin{equation}
{\cal P}_{\cal R} = A\ns{s}\left(\frac{k}{k_0}\right)^{n\ns{s}-1}\,,
\label{primordial_spectrum}
\end{equation}
where we have chosen the pivot scale $k_0=0.05\,{\rm Mpc}^{-1}$. We have not
investigated alternative initial conditions such as isocurvature scalar
perturbations or tensor perturbations, and we set the running of the spectral
index $\dd n\ns{s}/\dd\ln{k}=0$.

We treat the amplitude of the power spectrum $A\ns{s}$ as a nuisance
parameter that we cannot constrain. Nevertheless we use a wide prior range
for $A\ns{s}$ to get the normalized power spectrum required by the Planck
{\tt CamSpec} likelihood code. We configure the CLASS code to get the lensed
temperature power spectrum with reionization turned on for the timescape
matched FLRW model. We assume the reionization optical depth $\tau\ns{rion}$
to be completely degenerate with $A\ns{s}$.

\subsection{Choice of matched FLRW models}

Many of the matching methods of Sec.~\ref{mEH} require a shift
(\ref{Durrer_approx_2}) of very large $\ell$ multipoles of equivalent FLRW
models with large spatial curvatures of order $0.2\lsim\hOmkn\lsim0.8$. In
these cases the amount of time required to solve the Boltzmann code becomes
prohibitively large in combination with the MCMC analysis. We checked that
individual runs of the \Ar, \Ae, \At\ and \Ao\ matched models gave similar
$\chi^2$ values to the canonical \AH\ matched model, but were unable to
determine best fit parameter values with the computing resources available.
Our MCMC analysis is therefore restricted to matched models \AH, \Ww\ and \Wv.

The wall geometry matching methods \Ww\ and \Wv\ cannot give as good a
fit as the canonical \AH\ method to the volume--average expansion history.
However, they may provide better matching for those aspects of the
power spectra which are independent of the distance to the surface of last
scattering. Consequently, differences between the two methodologies also
provide a measure of the systematic uncertainties inherent in our matched
FLRW model approach.

\subsection{Foreground modeling}

Foregrounds can bias estimation of cosmological parameters, a problem which
is not unique to our study but also besets all analyses of CMB data sets in
the context of FLRW models as well. The {\tt CamSpec} likelihood in the
$50\leq\ell\leq 2500$ multipole range takes 14 additional parameters
that are used for relative calibration and unresolved foreground modeling. The
Planck team\footnote{See Sec.~4.2 and Appendix~C in \cite{Pparm}.} find that
foreground modeling does not change the parameter constraints in their
baseline six-parameter \LCDM\ model $\{\Omega_{\rm b}h^2$, $\Omega_{\rm c}h^2$,
$100\theta_{\rm MC}$, $\tau$, $n\ns{s}$, $\ln(10^{10}A_{\rm s})\}$.
However, they find that extensions to their baseline model are sensitive to
foreground modeling with the independent likelihood code {\tt Plik} converging
to slightly different values of $Y_{\rm p}$, $\Neff$, $n\ns{s}$. They
report up to a 1$\si$ shift in parameter values from the two likelihood codes
in some cases, but manage to obtain better agreement with the inclusion of
high--$\ell$ data which constrain the foregrounds to a higher precision.

We have investigated the possibility that foreground modeling may critically
impact the timescape parameter constraints. As a test we fixed the foreground
nuisance parameters to the best fit baseline \LCDM\ model values from Planck
\cite{Pparm}, with the hypothesis that the baseline \LCDM\ model unambiguously
identifies the foreground sources and their impact on the temperature
power spectrum. Fixing the foreground parameters in this way we found
the best fit parameters for the timescape cosmology to be
$\fvn=0.607^{+0.051}_{-0.057}$,
$\Hn=59.72^{+2.76}_{-2.89}\kmsMpc$, $10^{10}\etBg=6.24^{+0.3}_{-0.29}$,
$n\ns{s}=0.971^{+0.028}_{-0.030}$ (using the canonical \AH\ matched model).
Thus the chains converged to roughly the same region in parameter space, the
best fit parameters being within the 1$\si$ constraints from Table
\ref{tb:TS_params_val_Neff_fixed}, but with a larger value of
$-\ln{\cal L}=\chi^2/2$ as compared with the case in which foreground model
nuisance parameters are free to vary. These results are therefore consistent
with the hypothesis that foregrounds do not have a critical impact on
parameter estimation.

The precise value of the goodness of fit, $-\ln{\cal L}=\chi^2/2$, obtained
with fixed foreground nuisance parameters is not of acute importance since
we only use the Planck data, and do not combine it with other data sets.
If we were to combine the Planck data with other data sets that constrain
the CMB anisotropy spectrum, the issues around foreground modeling
would need to be revisited.

While we cannot entirely rule out the impact of foreground
modeling on timescape parameters we proceed with the assumption that our
results are not critically impacted by them. We treat the 14 foreground
parameters of the baseline model as nuisance parameters and
adopt the same prior range for these parameters as for the baseline
\LCDM\ model in the analysis by the Planck team \cite{Pparm}.

%-------------------------------------------------%
\begin{table*}[!htb]%htbp
\centering
\caption{\small\label{tb:TS_params_val_Neff_fixed}
The best fit and mean timescape parameter values,
with 68\% uncertainties. A fixed $\Neff=3.046$ is assumed.
The timescape models are matched to FLRW models with the same expansion history
until recombination up to density parameter differences of $<10^{-4}$. We
show only those matching methods for which the matched FLRW curvature
parameter is small enough to permit a full MCMC analysis using available
computer resources. The {\pAH}\ average expansion history matching method
provides the canonical constraint on timescape parameters. The wall expansion
history matching methods, {\protect\Ww}\ and {\protect\Wv}, are computed for
illustrative purposes.}
\makebox[\textwidth]{
\begin{tabular}{ @{}l c c c c c c }
\noalign{\vskip 3pt \hrule height 1.pt \vskip 3pt }
\multicolumn{1}{@{}l}{Matching type} &
\multicolumn{2}{l}{\hbox to50mm{\hfil\AH\hfil}%FLRW $k\neq 0$, $\Lambda>0$
} &
\multicolumn{2}{l}{\hbox to50mm{\hfil\Ww\hfil}%FLRW $k = 0$, $\Lambda =0$
} &
\multicolumn{2}{l}{\hbox to50mm{\hfil\Wv\hfil}%FLRW $k \neq 0$, $\Lambda =0$
} \\
\noalign{\vskip 3pt \hrule height 0.2pt\vskip 3pt }
Parameter &\hfil Best fit\hfil & Mean (68\% limits) \hfil &\hfil Best fit\hfil & Mean (68\% limits)\hfil& \hfil Best fit\hfil & Mean (68\% limits)\hfil\\
\noalign{\vskip 3pt\hrule height 0.2pt\vskip 3pt}
$\fvn$\dotfill & $0.628$ & $0.627^{+0.014}_{-0.012}$ & $0.545$ & $0.550^{+0.017}_{-0.015}$ & $0.559$ & $0.557^{+0.017}_{-0.015} $\\[5pt]

$\Hn$\dotfill & $60.984$ & $60.997^{+0.791}_{-0.733}$ & $56.148$ & $56.364^{+0.700}_{-0.694}$ & $56.774$ & $56.710^{+0.707}_{-0.714}$\\[5pt]

$10^{10}\etBg$\dotfill & $6.465$ & $6.489^{+0.101}_{-0.102}$ & $6.043$ & $6.048^{+0.086}_{-0.087}$ & $6.080$ & $6.080^{+0.091}_{-0.089}$\\[5pt]

$n\ns{s}$\dotfill & $0.995$ & $0.992 \pm 0.009$ & $0.956$ & $0.957 \pm 0.009$ & $0.963$ & $0.960 \pm 0.009$\\
\noalign{\vskip 3pt\hrule height 0.2pt\vskip 3pt}
$\Hb$\dotfill & $51.640$ & $51.658^{+0.251}_{-0.257}$ & $49.846$ & $49.882^{+0.223}_{-0.218}$ & $50.016$ & $49.996^{+0.227}_{-0.225}$\\[5pt]

$\Tb_0$\dotfill & $2.074$ & $2.074 \pm 0.010$ & $2.142$ & $2.137 \pm 0.013$ & $2.130$ & $2.131 \pm 0.013$\\[5pt]

$t_0$\dotfill & $16.587$ & $16.579 \pm 0.042$ & $16.640$ & $16.664^{+0.056}_{-0.054}$ & $16.675$ & $16.671^{+0.058}_{-0.053} $\\[5pt]

$\gbn$\dotfill & $1.314$ & $1.314 \pm 0.007$ & $1.273$ & $1.275 \pm 0.008$ & $1.280$ & $1.279 \pm 0.008$\\[5pt]

$Y_{\rm p}$\dotfill & $0.248$ & $0.248 \pm 0.00015$ & $0.247$ & $0.247 \pm 0.00014$ & $0.247$ & $0.247 \pm 0.00014$\\[5pt]

$\OMBn$\dotfill & $0.039$ & $0.039 \pm 0.001$ & $0.043$ & $0.043 \pm 0.001$ & $0.042$ & $0.043 \pm 0.001$\\[5pt]

$\OMCn$\dotfill & $0.176$ & $0.176^{+0.008}_{-0.010}$ & $0.238$ & $0.234^{+0.012}_{-0.013}$ & $0.227$ & $0.228^{+0.012}_{-0.013}$\\[5pt]

$\OMMn$\dotfill & $0.215$ & $0.216^{+0.009}_{-0.011}$ & $0.281$ & $0.277^{+0.012}_{-0.014}$ & $0.269$ & $0.271^{+0.012}_{-0.014}$\\[5pt]

$\OMkn$\dotfill & $0.819$ & $0.818^{+0.010}_{-0.008}$ & $0.757$ & $0.762^{+0.013}_{-0.012}$ & $0.768$ & $0.767^{+0.013}_{-0.011}$\\[5pt]

$\OM_{{\cal Q}0}$\dotfill & $-0.034$ & $-0.034 \pm 0.001$ & $-0.038$ & $-0.038 \pm 0.001$ & $-0.038$ & $-0.038 \pm 0.001$\\[5pt]

$\bar z\ns{dec}$\dotfill & $1429.729$ & $1429.425^{+7.171}_{-6.165}$ & $1387.674$ & $1390.524^{+8.570}_{-7.723}$ & $1394.902$ & $1394.064^{+8.473}_{-7.723}$\\[5pt]

$\bar z\ns{drag}$\dotfill & $1396.292$ & $1396.341^{+8.081}_{-6.996}$ & $1348.949$ & $1351.776^{+9.231}_{-8.309}$ & $1356.537$ & $1355.719^{+8.522}_{-8.515}$\\[5pt]

$\bD_{\rm s}(\bar z\ns{dec})$\dotfill & $0.135$ & $0.135 \pm 0.001$ & $0.132$ & $0.132 \pm 0.001$ & $0.133$ & $0.133 \pm 0.001$\\[5pt]

$\bD_{\rm s}(\bar z\ns{drag})$\dotfill & $0.140$ & $0.140 \pm 0.001$ & $0.139$ & $0.139 \pm 0.001$ & $0.139$ & $0.139 \pm 0.001$\\
\noalign{\vskip 3pt\hrule height 0.2pt \vskip 3pt}
$\tau_{\rm w0}$\dotfill & $13.811$ & $13.806^{+0.067}_{-0.065}$ & $14.267$ & $14.261^{+0.058}_{-0.052}$ & $14.229$ & $14.233^{+0.059}_{-0.054}$\\[5pt]

$\OmBn$\dotfill & $0.089$ & $0.089 \pm 0.001$ & $0.089$ & $0.089 \pm 0.001$ & $0.089$ & $0.089 \pm 0.001$\\[5pt]

$\OmCn$\dotfill & $0.400$ & $0.400^{+0.014}_{-0.016}$ & $0.490$ & $0.484 \pm 0.017$ & $0.475$ & $0.477 \pm 0.017$\\[5pt]

$\OmMn$\dotfill & $0.489$ & $0.489^{+0.014}_{-0.016}$ & $0.579$ & $0.573 \pm 0.017$ & $0.564$ & $0.566 \pm 0.017$\\[5pt]

$z\ns{dec}$\dotfill & $1087.568$ & $1087.503^{+0.480}_{-0.520}$ & $1090.141$ & $1090.039^{+0.494}_{-0.495}$ & $1089.800$ & $1089.822^{+0.498}_{-0.507}$\\[5pt]

$z\ns{drag}$\dotfill & $1062.128$ & $1062.323^{+0.723}_{-0.732}$ & $1059.713$ & $1059.655^{+0.638}_{-0.646}$ & $1059.820$ & $1059.836^{+0.666}_{-0.667}$\\[5pt]

$100\,\theta_{\rm dec}$\dotfill & $1.048$ & $1.047 \pm 0.001$ & $1.041$ & $1.041 \pm 0.001$ & $1.041$ & $1.041 \pm 0.001$\\[5pt]

$100\,\theta_{\rm drag}$\dotfill & $1.064$ & $1.064 \pm 0.001$ & $1.061$ & $1.061 \pm 0.001$ & $1.061$ & $1.061 \pm 0.001$\\[5pt]

$d_{\rm A}$\dotfill & $12.888$ & $12.883 \pm 0.063$ & $12.699$ & $12.729^{+0.067}_{-0.066}$ & $12.753$ & $12.750^{+0.067}_{-0.066}$\\[5pt]

$d_{\mathrm{A,drag}}$\dotfill & $13.191$ & $13.183 \pm 0.062$ & $13.058$ & $13.088^{+0.066}_{-0.065}$ & $13.108$ & $13.104^{+0.067}_{-0.066}$\\
\noalign{\vskip 3pt \hrule height 1.pt \vskip 3pt}
\end{tabular}}
\end{table*}
%-------------------------------------------------%

\section{Results and analysis}\label{results}

Here we present the parameter constraints obtained from the Planck temperature
power spectrum data in the range $50\leq\ell\leq 2500$. Our analysis includes
the parameters that model the foreground in this multipole range \cite{Pparm}.

In Table \ref{tb:TS_params_val_Neff_fixed} we record the best fit and mean
marginalized constraints on timescape parameters obtained for each of the
matching procedures \AH, \Ww\ and \Wv\ with $\Neff=3.046$ fixed.
The chains analyses and all statistical information were obtained
with the {\tt getdist} software in the CosmoMc package \cite{Lewis_&_Bridle}.
We ran eight chains for the canonical \AH\ matching method and five chains each
for the \Ww\ and \Wv\ methods. We have checked to ensure that the base MCMC
parameters satisfy the Gelman--Rubin\footnote{The Gelman--Rubin diagnostic is
used to test that all chains converge to the same posterior distribution
\cite{Gelman_&_Rubin}. The $R-1$ statistic in the diagnostic compares the
variance of the chain means to the mean value of the variances within each
chain. Generally $R-1<0.2$ is optimal but even then low values indicate, but
do not guarantee, convergence.} diagnostic $R-1<0.01$ and that the chains have
converged. Not all of the derived parameters satisfy these criteria but even
then all parameters satisfy $R-1<0.02$ and are in fact closer to $0.01$ than
to $0.02$.

In what follows we treat the \AH\ model as the canonical
matching procedure and its results will be used for timescape parameter
constraints. The other matching procedures are used for the purpose
of understanding the effects of matching assumptions on parameter constraints
and possible systematic uncertainties.
%-------------------------------------------------%
\begin{figure*}[t]
\begin{center}
\includegraphics[height=14cm,width=16cm]{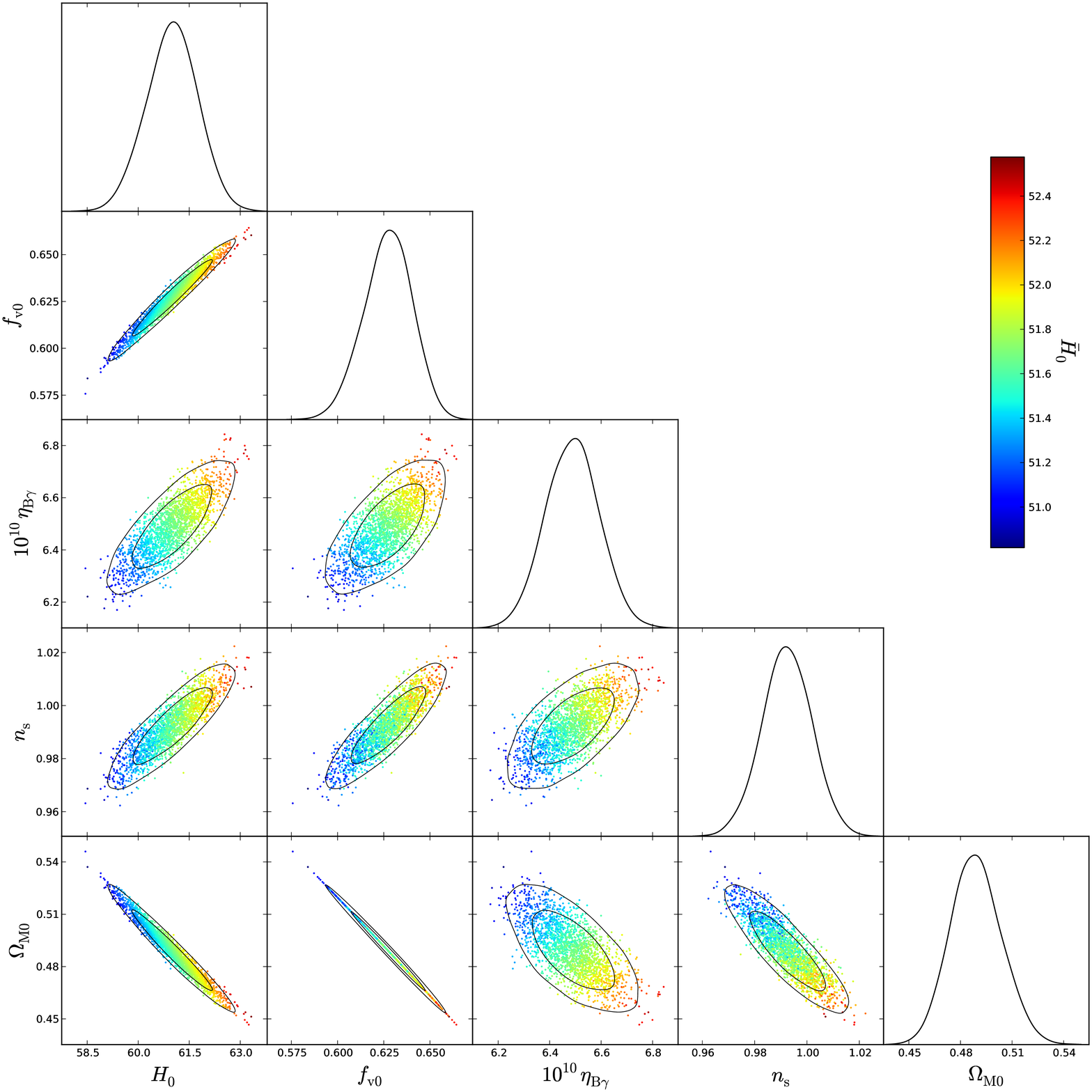}
\end{center}
\caption{Constraints on the base MCMC parameters for the timescape parameters
listed in Table \ref{tb:TS_params_def}, and the total dressed matter density,
$\OmMn$ %$\OM$ relevant to a wall/galaxy observer. The parameter constraints
for timescape parameters with matched model \pAH, showing 68\% and 95\%
statistical uncertainties. We also show a scatter plot of the parameters colour
coded with the value of the bare Hubble constant, $\protect\Hb$, relevant to
a volume--average expansion.}
\label{fig:triangle}
\end{figure*}
%-------------------------------------------------%

In Fig.~\ref{fig:triangle} the marginalized probability distribution and
2--dimensional posterior constraints for the base timescape parameters and the
dressed total matter density are shown for the canonical \AH\ matching. The
constraints on all parameters can be understood in terms of the angular scale
of the sound horizon
$\th\ns{dec}=\bD_{\rm s}(\bar z\ns{dec})/d_{\rm A}$
and the ratios of the acoustic peaks.

Before proceeding any further let us first make the relationship between
$\OmMn$ and $\fvn$ conspicuous. At late epochs the full timescape solution
with radiation is very well approximated by the tracker limit of the matter
only solution \cite{sol}, leading to the following relations between
the present void fraction and dressed matter density
\beq
\OmMn\approx\frn12\left(1-\fvn \right)\left(2+\fvn \right)\,,\quad\left.
\frac{\dd\Omega\ns{M}}{\dd\fv}\right|_0\approx-\left(\frn12+\fvn\right),
\label{eq:OmegaM_from_fv0_tracker}
\eeq
which explains the negative correlation between $\OmMn$ and $\fvn$. The Planck
data tightly constrain the angular position of the first peak, $\th\ns{dec}$,
and therefore any changes in $\Hn$, $\OmMn$, $\fvn$ which lead to an increase
in $d_{\rm A}$ simultaneously result in an increase in $\bD_{\rm s}$ and
vice versa. A larger angular diameter distance, $\dA$, to the last scattering
surface can be obtained by increasing either $\Hn$ or $\fvn$ but there is a
concurrent increase in $\bD_{\rm s}$, which also changes the ratios of
the acoustic peak heights.

Matching the ratios of the peak heights requires a delicate balance between the
proportions of baryonic and nonbaryonic matter. The extent to which $\fvn$ can
change while keeping $\th\ns{dec}$ constant is limited because the ratio
$\OmMn/\OmBn$ (or equivalently $\fvn/\etBg$) fixes the ratios of the first,
second and third acoustic peak heights, which are also tightly constrained by
the Planck data.

In the FLRW models one of the two parameters, $\Omkn$ and $\OmLn$ (as
constrained from the Friedman equation) can be adjusted to determine the
overall angular scale of the peaks without directly influencing the ratios of
the acoustic peak heights. By contrast, in the timescape model the void
fraction constrains both the matter density parameter -- either dressed
(\ref{eq:OmegaM_from_fv0_tracker}) or bare $\OMMn\approx4(1-\fvn)/(2+\fvn)^2$
-- and the volume average curvature density parameter, $\OMkn\approx 9\fvn/
\left(2+\fvn\right)^2$. Thus timescape parameters are very tightly constrained;
a change to $\fvn$ can significantly affect both the angular scale and,
insofar as it changes the ratio $\OmBn/\OmMn$, also the ratios of the peak
heights.

The results of MCMC analysis on the canonical \AH\ model yield $H_0=61.0^
{+0.79}_{-0.73}\kmsMpc$ and $\fvn=0.627^{+0.014}_{-0.012}$. The value
of the dressed Hubble constant agrees with our previous estimate, $61.7\pm
3.0\kmsMpc$ \cite{dnw} at the 1$\si$ level. However, the void fraction
is almost 2$\si$ less than the previous estimate \cite{dnw}, $\fvn=0.695^
{+0.041}_{-0.051}$. Of course, we should caution that the uncertainties
given in Table \ref{tb:TS_params_val_Neff_fixed} do not include any estimate of
the systematic uncertainties that must surely arise from the matched FLRW
procedure. Since the wall geometry based matching procedures \Ww\ and \Wv\ do
not reliably estimate volume average quantities\footnote{The present epoch
values of the void fraction for the \Ww\ and \Wv\ methods listed in Table
\ref{tb:TS_params_val_Neff_fixed} have $\fvn<0.587$ and thus represent models
with no apparent cosmic acceleration \cite{clocks,sol}, whereas the \AH\
matched model does have apparent acceleration},
the significantly smaller values of $\Hn$, $\fvn$ obtained for these
procedures probably overestimate the systematic uncertainties. However,
as an upper bound they indicate that the systematic uncertainties could
be as high as 8--13\%, as compared to the 1--2\% statistical uncertainties.

Even if the systematic uncertainties were not so large, however, there is
an obvious reason for the apparent tension between
the two estimates of $\fvn$. In all previous work \cite{clocks,lnw,dnw}
we have not directly constrained the baryon--to--photon ratio. In fact,
our best previous estimate \cite{dnw} is based on assuming a
baryon--to--photon ratio $10^{10}\etBg=5.1\pm0.5$ for which one can avoid
a primordial lithium abundance anomaly \cite{bbn,pill}. By contrast, here
we have used the acoustic peaks height ratio to directly constrain $\etBg$
for the first time, with the result $10^{10}\etBg = 6.49\pm 0.10$. If we
had admitted such large values of the baryon--to--photon ratio in our
previous estimate then there would not be a discrepancy.

We find that while the timescape model remains observationally consistent a
detailed analysis of the acoustic peaks in the Planck data -- modulo
systematic uncertainties introduced by the model matching procedure -- the
baryon--to--photon ratio is driven to a value that is even higher
than \LCDM\ model estimates \cite{Pparm,wmap9}. Thus based on the analysis
here we cannot make the claim that the timescape model solves the primordial
lithium abundance anomaly.

One might be concerned that the value of the baryon--to--photon ratio is
4$\,\si$ larger than the \LCDM\ value $10^{10}\etBg=6.04\pm0.09$. However, the
wall geometry based matching procedures give values $10^{10}\etBg=6.05\pm0.09$
and $10^{10}\etBg=6.08\pm0.09$, which precisely match the \LCDM\ result.
Considering the systematic uncertainties therefore, we have agreement with
\LCDM.

The spectral index is also constrained for the first time. The canonical
\AH\ matching method leads to a nearly scale invariant primordial spectrum
with $n\ns{s} = 0.992 \pm 0.009$, $n\ns{s}=0.960\pm0.009$ whereas the
wall--geometry matching methods yield $n\ns{s}=0.957\pm0.009$ and
$n\ns{s}=0.960\pm0.009$, confirming deviations from scale invariance at more
than $3\,\si$ level in agreement with the \LCDM\ results \cite{Pparm,wmap9}.
Once again, these 3--4\% differences are driven by the systematic uncertainties
that arise from the imperfect nature of the matched model procedures.

Since $10^{10}\etBg$ and $n\ns{s}$ constrain spectral features other than the
overall angular scale, the difference between the average--geometry and
wall--geometry matching procedures give a reasonable estimate of the
systematic uncertainties. In fact, the the wall--geometry matching
procedures produce a somewhat reduced value of $-\ln{\cal L}=\chi^2/2$
as compared to the volume--average methods, with $-\ln{\cal L}=3925.16$,
$3897.90$ and $3896.47$ for the \AH, \Ww\ and \Wv\ methods respectively.
The likelihoods for the \Ww, \Wv\ matching methods are in fact precisely
the same as one obtains for best fit \LCDM\ model in the multipole range
$50\leq\ell\leq 2500$, where the Planck team obtain \cite{prod}
$-\ln{\cal L}=3895.5$ using {\tt MINUIT} or $-\ln{\cal L}=3896.9$ using {\tt
CosmoMC}.

It therefore appears that the FLRW perturbation theory for the wall--geometry
matching methods produces a better fit to the features of the acoustic peaks
which relate solely to its shape. It is possible that the values of $10^{10}
\etBg$ and $n\ns{s}$ obtained by wall geometry matching therefore give a more
accurate estimate of the values that we would obtain if we could both use the
most relevant perturbation equations, and simultaneously use the timescape
solution in all of the codes, rather than having to rely on matched models
using CLASS \cite{class_I,class_II}.
%-------------------------------------------------%
\begin{figure}[!htb]
\begin{center}
\includegraphics[width=8cm,height=6.5cm]{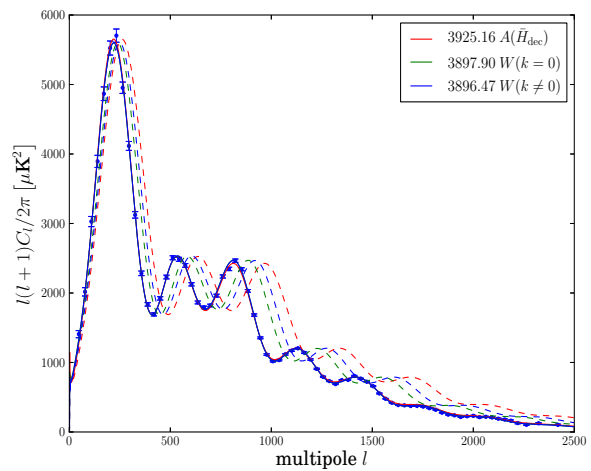}
\end{center}
\caption{\small Angular power spectra for the best fit parameters of the three
matching methods \pAH\ (red); \Ww\ (green); \Wv\ (blue), with $-\ln{\cal L}$
values shown. In each case the dashed lines show the spectrum before the
application of the shift (\ref{Durrer_approx_2}), and the solid lines after.
Data points, plotted for $\ell\ge50$, are derived from the {\tt CAMspec}
likelihood \cite{Pparm}, with $1\,\si$ error bars including beam and foreground
uncertainties.}
\label{fig:Cls}
\end{figure}
%-------------------------------------------------%
\begin{figure}[!htb]
\begin{center}
\includegraphics[width=8cm,height=6.5cm]{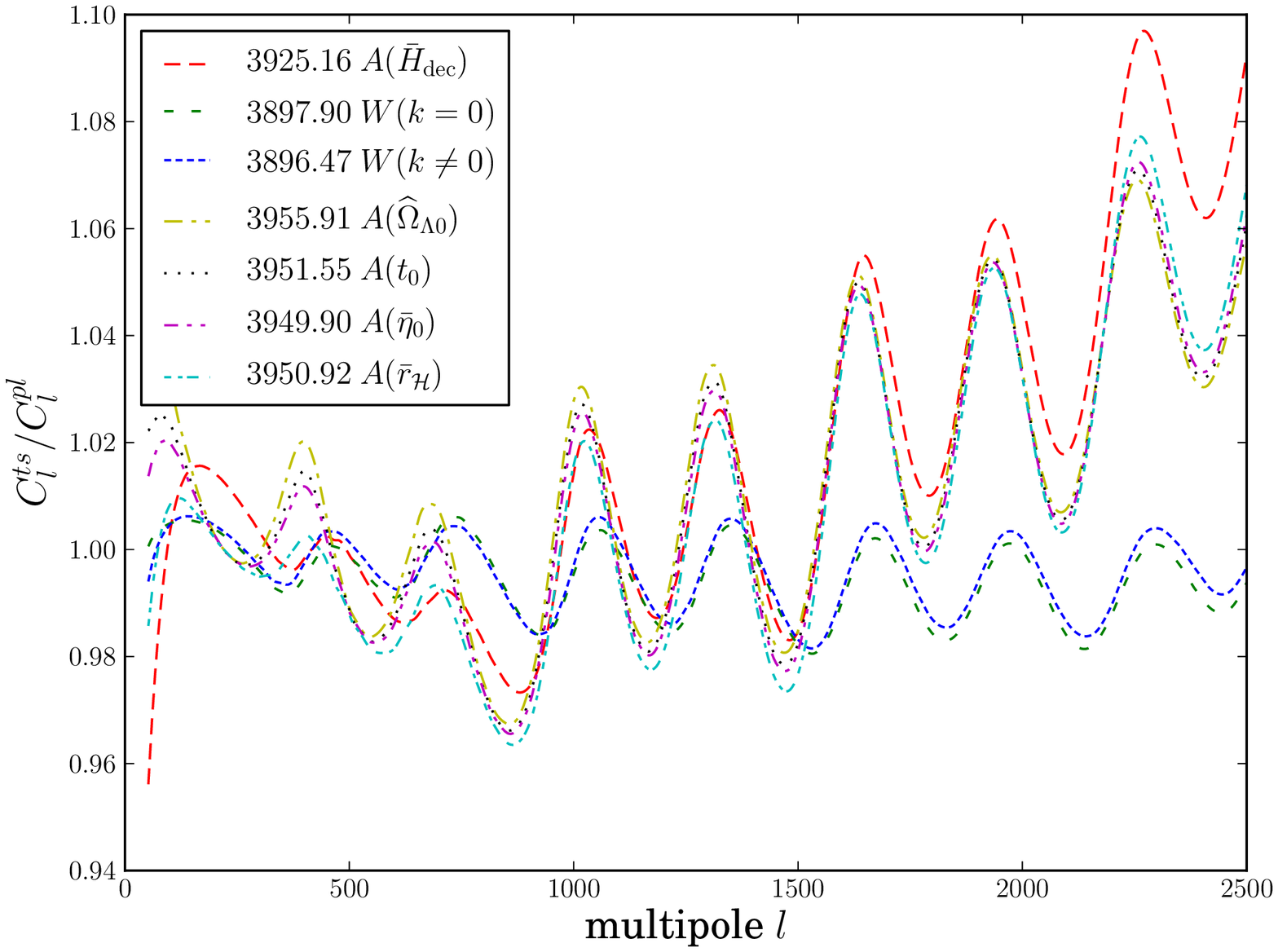}
\end{center}
\caption{\small Ratios of multipoles of matched timescape models to those of
the \LCDM\ model with the best fit Planck parameters, which is used as a
reference model. For the \pAH, \Ww\ and \Wv\ methods the best fit values from
Table~\ref{tb:TS_params_val_Neff_fixed} have been used. For the {\protect\Ar},
{\protect\Ae}, \At\ and \Ao\ we have used the best fit parameters from the
\pAH\ model. The values of $-\ln{\cal L}=\chi^2/2$ shown do not therefore
represent the best fit values in these cases.}
\label{fig:Cls_ratios}
\end{figure}
%-------------------------------------------------%
\begin{figure*}[!t]
\begin{center}
\includegraphics[width=15cm,height=15cm]{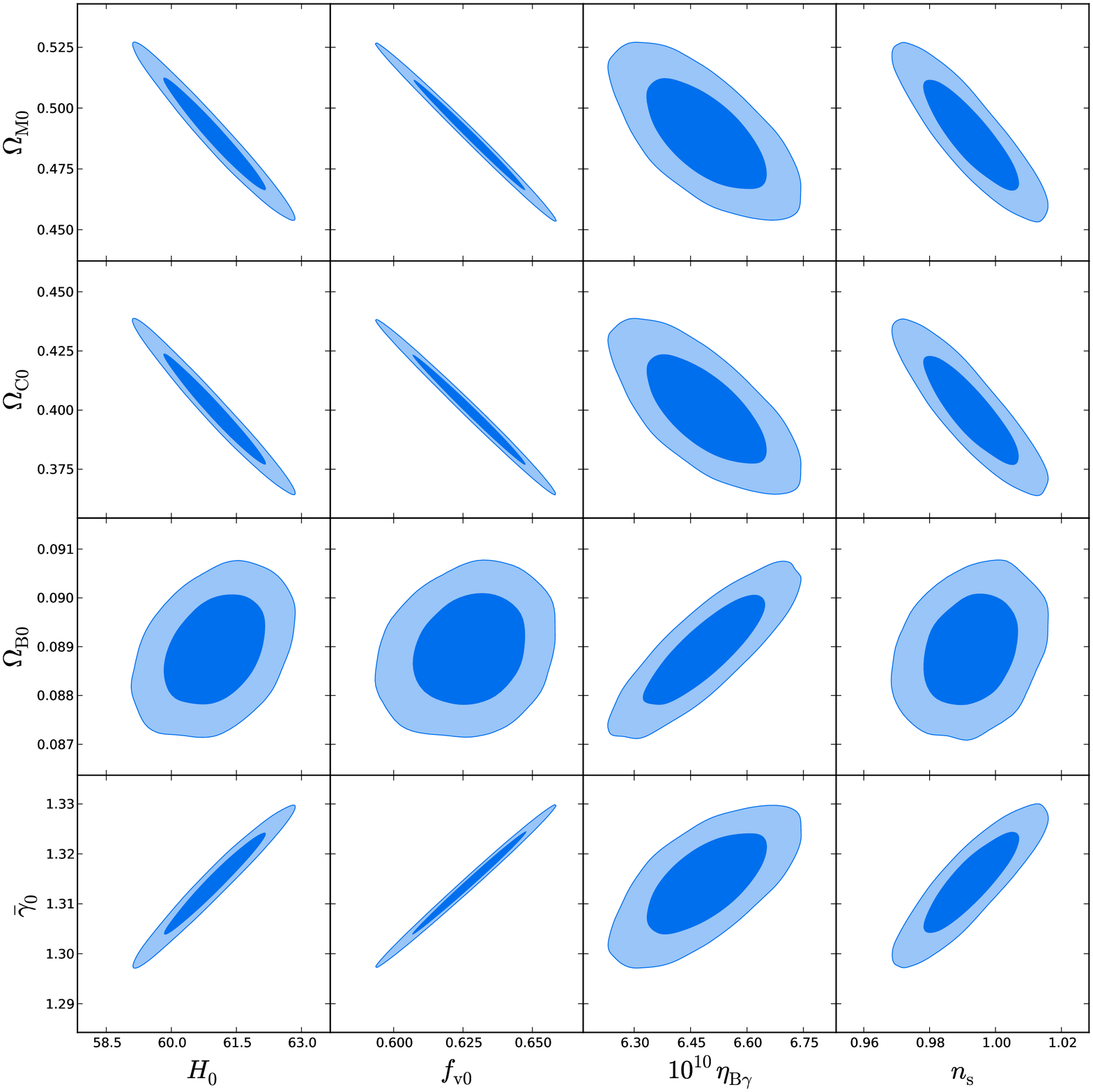}
\end{center}
\caption{\small Variations in the various dressed parameters are shown with
respect to the base MCMC parameters for the timescape model. The parameter
constraints -- showing 68\% and 95\% statistical uncertainties -- are
determined for the case of the \pAH\ matching method.}
\label{fig:probs_chains_ts_lcdm_rect_dressed}
\end{figure*}
%-------------------------------------------------%
\begin{figure*}[!p]
\begin{center}
\includegraphics[width=15cm,height=20cm]{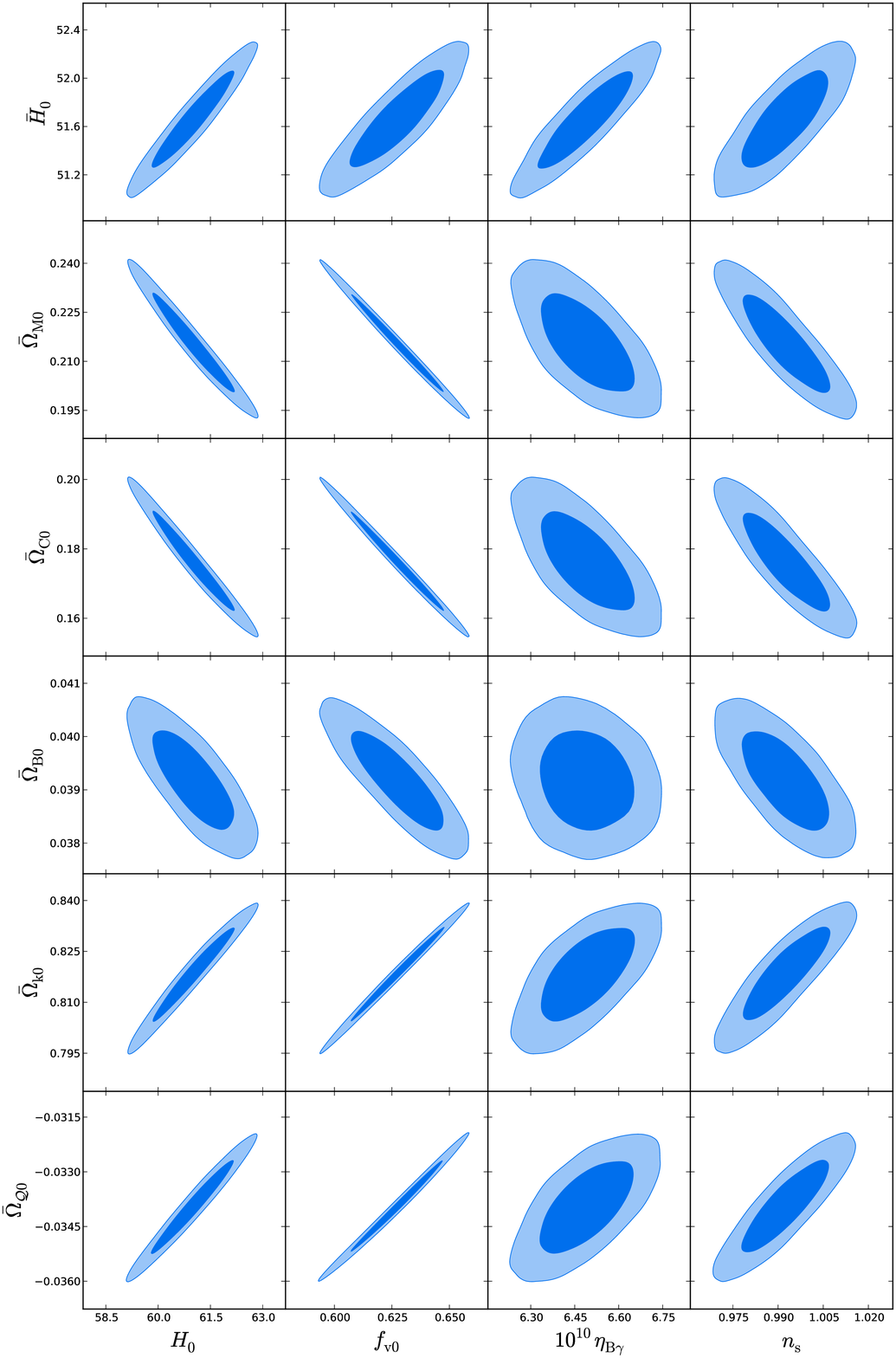}
\end{center}
\caption{\small Variations in the various bare parameters are shown with
respect to the base MCMC parameters for the timescape model. The parameter
constraints -- showing 68\% and 95\% statistical uncertainties -- are
determined for the case of the \pAH\ matching method.}
\label{fig:probs_chains_ts_lcdm_rect_bare}
\end{figure*}
%-------------------------------------------------%

In Fig.~\ref{fig:Cls} the power spectra for the best fit values
are shown for the three matching procedures for which the MCMC analysis
was possible.
To show the level of concurrence of the other volume--average expansion
history matching procedures from Sec.~\ref{mEH}, we determined their
power spectra using the best fit parameters for
\AH\ matching found by the MCMC analysis. In Fig.~\ref{fig:Cls_ratios}
we plot the ratios of these power spectra relative to a fiducial \LCDM\
spectrum obtained from the best fit parameters using the Planck data only
\cite{Pparm}. We also plot the same ratio for the three best fit models
of Table~\ref{tb:TS_params_val_Neff_fixed}.

We see in Fig.~\ref{fig:Cls_ratios} that the difference in individual $C_\ell$
values from the \LCDM\ model is of up to order 1\% for the wall expansion
history matching methods over all multipoles. For the volume--average matching
methods the differences are individually up to 2--3\% for $50\lsim\ell\lsim
1600$, and slowly rise to a maximum $\goesas8$--9\% for $2300\lsim\ell\lsim
2500$. While the small angle differences may seem large, it must be remembered
that individual foreground parameters will also be somewhat different so that
the overall $(-\ln{\cal L})$ value for the best fit \AH\ matched model is only
0.7\% larger than for the best fit \LCDM\ model \cite{Pparm}.

We note that in Fig.~\ref{fig:Cls_ratios} the increased power in the canonical
\AH\ matching method as compared to the fiducial \LCDM\ power spectrum at very
large multipoles will be partly due to the difference in spectral index
$n\ns{s}=0.992\pm0.009$ compared to $n\ns{s}=0.9616\pm0.0094$. Increasing
$n\ns{s}$ increases the power at the third peak and higher multipoles (small
angles) compared to the first two peaks \cite{Mtext}.

In Figs.~\ref{fig:probs_chains_ts_lcdm_rect_dressed} and
\ref{fig:probs_chains_ts_lcdm_rect_bare} the posterior constraints on the
dressed and bare parameters are shown against the base MCMC parameters.
These figures illustrate a number of interesting points relating to the
constraints $\etBg$ and $n\ns{s}$.
In Fig.~\ref{fig:probs_chains_ts_lcdm_rect_bare} the bare baryon density
$\OMBn$ appears to be uncorrelated to $\etBg$. Thus since $\etBg\propto\OMBn
\Hb^2$, any increase in $\etBg$ must be met with an increase in $\Hb$. By
contrast, the dressed baryon density parameter, $\OmBn$ is positively correlated
with $\etBg$, as shown in Fig.~\ref{fig:probs_chains_ts_lcdm_rect_dressed}.
The phenomenological lapse function, $\gb$, and its time derivative, $d\gb/dt$,
which relate the dressed and bare parameters according to $\OmBn=\OMBn/\gbn^3$
and $\Hn=\left.(\gb\bH-d\gb/dt)\right|\Z0$, are involved here in
subtle ways which are difficult to disentangle. This also explains
why the volume--average geometry and wall--geometry matching methods
lead to different results for $\etBg$. Essentially, one cannot separate
parameters concerning the average expansion history from those that
relate to spectral features such as the ratios of acoustic peak heights.

The variation in the spectral index, $n\ns{s}$, between the
different matching methods can be similarly understood.
Figures \ref{fig:probs_chains_ts_lcdm_rect_dressed} and
\ref{fig:probs_chains_ts_lcdm_rect_bare} reveal a degeneracy between
$n\ns{s}$ and the dressed and bare parameters $\OmMn$, $\OMMn$, $\OmBn$,
$\Hb$. One cannot disassociate the early universe physics which is influenced
by $n\ns{s}$ from the late time evolution on account of parameter degeneracies.
The lesson here is that, as yet, the constraints on the timescape parameters
are limited by systematic uncertainties from the matching method which are
larger than the statistical uncertainties. Furthermore, the exclusion of
$\ell<50$ data is also a handicap for us since the Sachs--Wolfe plateau on
the largest scales is sensitive to $n\ns{s}$, where the largest wavelengths
remain frozen outside the horizon before recombination and are unaffected
by causal physics.

\begin{figure*}[!p]
\begin{center}
\includegraphics[width=15cm,height=20cm]{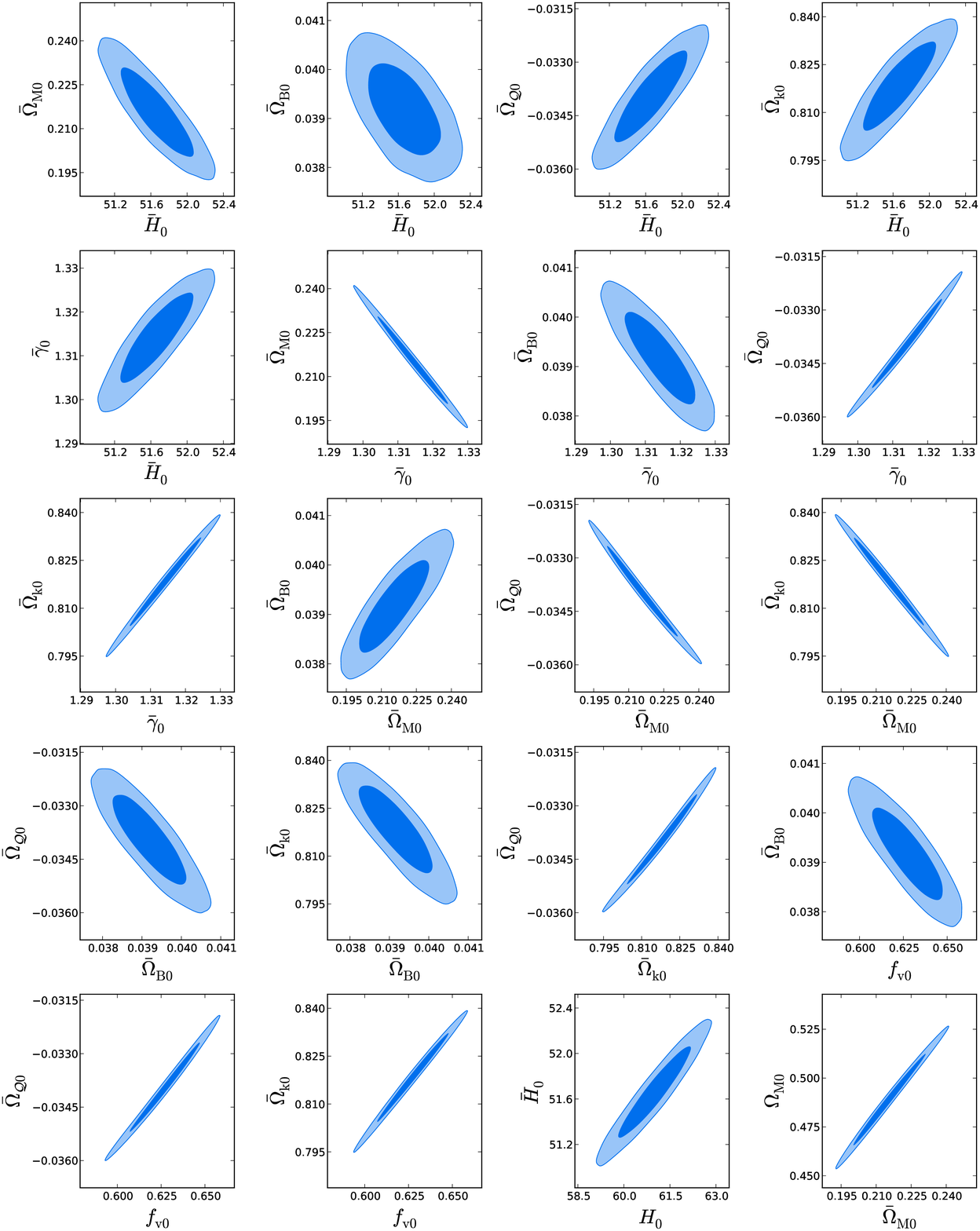}
\end{center}
\caption{\small Correlations between selected bare and dressed timescape
parameters are shown. The parameter constraints -- showing 68\% and 95\%
statistical uncertainties -- are determined for the case of
the \pAH\ matching method.}
\label{fig:probs_chains_ts_lcdm_2D}
\end{figure*}

In Fig.~\ref{fig:probs_chains_ts_lcdm_2D} the correlations between the derived
bare and dressed parameters are shown. The narrow 68\% and 95\% confidence
contours between various parameters largely reinforce what we already know
about their degeneracies from the timescape matter--only tracker solution
\cite{sol,obs}. In the limit of vanishing radiation energy density the width of
the contours would shrink to zero. The 2--dimensional posteriors involving
baryon density are not as tightly constrained because the tracker solution fixes
the total matter density but not its split into baryonic and nonbaryonic
components. Interpreting $\Hn$ as the Hubble constant determined
by wall observers, and $\Hb$ as the Hubble constant of the volume--average
statistical geometry, the contours in the $\Hn$ -- $\Hb$ plane in
Fig.~\ref{fig:probs_chains_ts_lcdm_2D} show that bounds on the bare parameter
provide reasonably tight bounds on the dressed parameter and vice versa.

\begin{figure*}[!p]
\begin{center}
\includegraphics[scale=0.4]{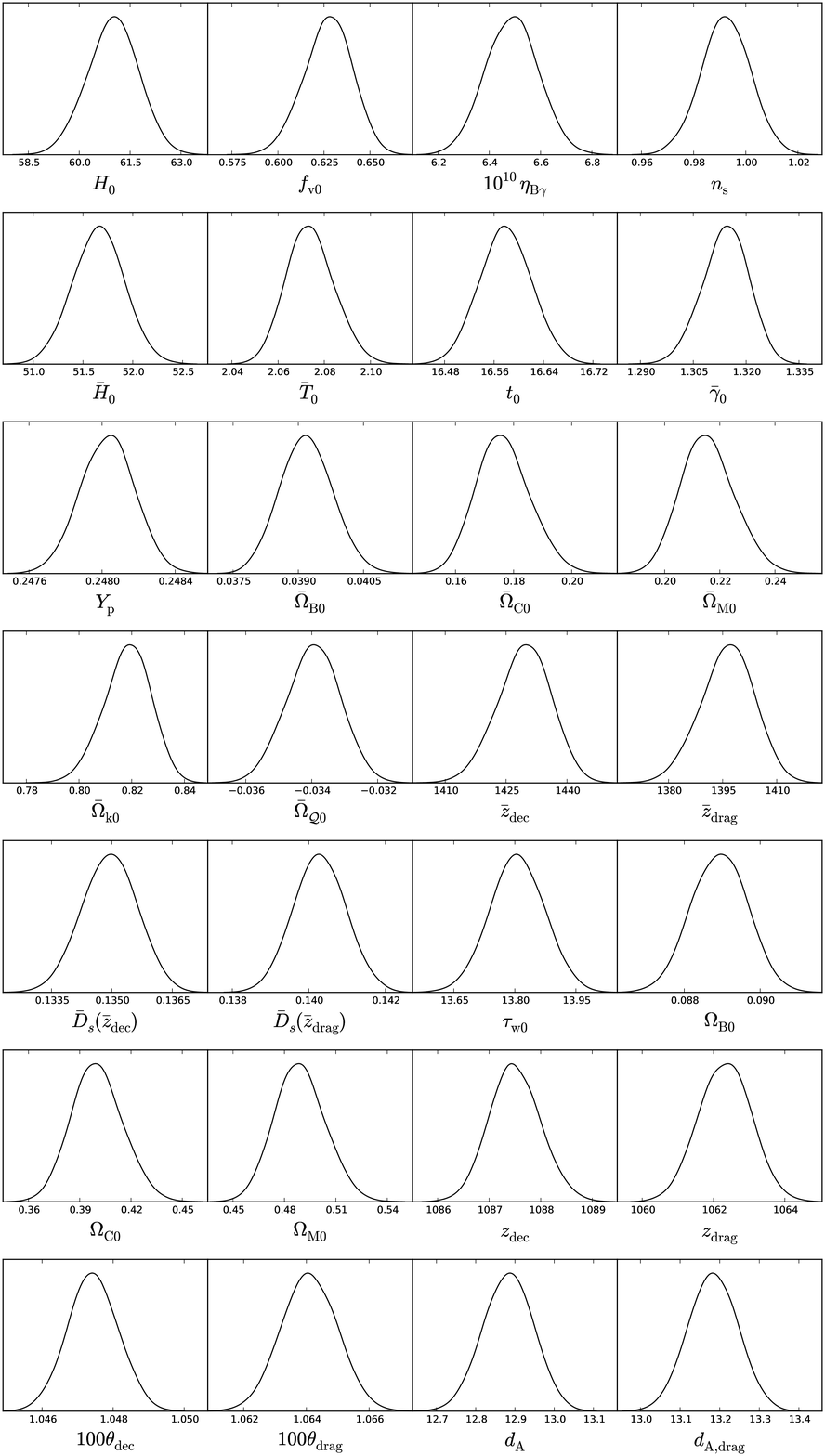}
\end{center}
\caption{\small Marginalized posteriors for the base MCMC parameters, the
bare parameters and the dressed parameters in the timescape cosmology.
The parameter constraints are determined for the case of the \pAH\
matching method.}
\label{fig:probs_chains_ts_lcdm_1d}
\end{figure*}

Finally, in Fig.~\ref{fig:probs_chains_ts_lcdm_1d} the marginalized likelihoods
for all the timescape parameters of the canonical matching listed in Table
\ref{tb:TS_params_def} are shown for completeness.

%-------------------------------------------------%
%-------------------------------------------------%
%-------------------------------------------------%

Among the derived parameters in Table~\ref{tb:TS_params_val_Neff_fixed},
it is interesting to note that the age of the Universe as measured by a
wall observer, $\tau_{\rm w0}=13.81\pm0.07\,$Gyr, matches the standard
\LCDM\ fit to Planck \cite{Pparm}. It should be stressed that this numerical
result is not at all related to the matched \LCDM\ model, which for the \AH\
method in the best fit case has $\hOmLn=0.7633$, $\hOmMn=0.2153$, $\hOmkn=
0.02138$ and a matched age of $19.81\,$Gyr in volume--average time.
Wall geometry matching gives $\tau_{\rm w0}=14.26\pm0.06\,$Gyr
and $\tau_{\rm w0}=14.23\pm0.06\,$Gyr for the \Ww\ and \Wv\ methods
respectively, once again indicating that the systematic uncertainties are
larger than the statistical uncertainties, being of order 3.5\% in this case.

The present comoving scale of the baryon acoustic oscillation,
which is given by
\beq
r\Ns{drag}=(1+z\Ns{drag})\th\Ns{drag}d\Ns{A,drag}
\eeq
is determined by the parameters of Table~\ref{tb:TS_params_val_Neff_fixed}
to be $r\ns{drag}=149.3\pm0.7\,$Mpc for the canonical \AH\ matched model,
and $r\ns{drag}=147.0\pm0.7\,$Mpc and $r\ns{drag}=147.5\pm0.7\,$Mpc for the
\Ww\ and \Wv\ methods respectively. The systematic uncertainties between
the different methods, of order 1.5\%, are smaller than for other parameters.
In this case the \Ww\ and \Wv\ values precisely match the \LCDM\ best fit value
from Planck, $r\ns{drag}=147.34\,$Mpc.

\section{Discussion}\label{dis}

We have for the first time investigated the parameter bounds that can
be put on the timescape cosmology through a full analysis of the acoustic
peaks in the power spectrum of CMB anisotropies, using Planck satellite data.
Since the expansion history of the timescape model differs only slightly from
that of a FLRW model at early epochs, we performed our analysis
by directly computing relevant processes (such as nucleosynthesis and
recombination) for the timescape model in the early Universe, and then matched
the expansion history of the timescape model to that of the closest equivalent
FLRW model in volume--average coordinates. We investigated a number of
matching procedures, but used the \AH\ matching to provide statistical bounds
on parameters related to the average expansion history.

We conclude that there are parameters for which the timescape model is
a good fit to the Planck data, and it remains competitive with the \LCDM\
model. However, we can no longer claim a resolution of the primordial lithium
abundance problem, as seemed possible with fits based solely on the
angular diameter distance of the sound horizon at decoupling and the baryon
drag epoch \cite{dnw}. This is our most important result.

It is clear, however, that this conclusion is driven by the ratio of the
heights of the acoustic peaks which depend strongly on the ratio
of baryons to nonbaryonic dark matter in the primordial plasma. Furthermore,
these results depend heavily on perturbation theory in the early Universe
for which a standard \LCDM\ model has been assumed, with possibly
different constituent ratios.

Although backreaction is negligible in determining the background solution,
in our numerical examples $\OM\Z{\QQ\dec}\goesas-1\times10^{-5}$ is of
the same order as the density perturbations, $\de\rho/\rho$, in baryons at
decoupling. Thus although the backreaction terms are inconsequential in
determining the background at early times they very probably should {\em not}
be neglected in considering the evolution of perturbations in determining the
acoustic peaks. By analogy, in the standard treatment both density and velocity
perturbations are small and do not significantly affect the background but
being of similar order they must be considered as a coupled system.
The fact that different matched FLRW model procedures
give rise to systematic uncertainties of 8--13\% in some present epoch
parameters, is a direct demonstration that differences of order $10^{-5}$
between the matched models at decoupling are nonetheless significant.

Ideally therefore we should consider the backreaction formalism with
pressure \cite{buch01} for a matter plus radiation plasma which begins
in close to homogeneous and isotropic state. Since the average evolution
is not {\em exactly} a Friedmann model, such an analysis is subtly different
from a perturbative approach which assumes that average evolution is
exactly a solution of Einstein's equations. The problem of backreaction
in the primordial plasma has not been studied in any more detail
than Buchert's initial formal study \cite{buch01}. There is much
discussion of backreaction in models which are close to FLRW backgrounds,
but debate has centred on the question of whether backreaction can cause
significant deviations of the average evolution from FLRW evolution in
the case of dust cosmologies \cite{CELU}. The question here differs both by the
virtue of the matter content, and by the fact that we are interested in changes
to the growth of perturbations rather than the average evolution itself.
As far as physical cosmology is concerned, this is completely uncharted
territory.

To summarize our results, if we take the canonical \AH\ matching method
to estimate the parameters which describe the average expansion history,
but use the difference in the wall geometry matching method results of
Table~\ref{tb:TS_params_val_Neff_fixed} to estimate systematic uncertainties
then for the volume--average base parameters, we have\footnote{In the case
of the systematic uncertainties, the lower bound is what was actually
obtained using the \Ww\ and \Wv\ matching procedures in
Table~\ref{tb:TS_params_val_Neff_fixed}. However, since other matching
procedures could be also envisioned, we use the bound obtained as a best
estimate percentage systematic uncertainty.} $\Hn=61.0\kmsMpc$
($\pm1.3$\% stat) ($\pm8$\% sys), and a present void volume fraction
$\fvn=0.627$ ($\pm2.33$\% stat) ($\pm13$\% sys). This corresponds to a
dressed matter density\footnote{While this parameter is numerically closer to
the FLRW value than the bare parameter is, since it does not obey a
Friedmann--like equation it is not directly subject to the observational
constraints that are placed on the parameter $\OmMn$ of the standard
cosmology. The two parameters cannot be equated.} $\OmMn=0.489$ ($\pm3.3$\%
stat) ($\pm20$\% stat), which is consistent with the bounds on this
parameter obtained from supernovae data using the SDSS-II
sample \cite{Kessler09}, as shown in Fig.~8 of Ref.\ \cite{sw} .

Since the wall--geometry matching methods are better adapted to those aspects of
the timescape model that do not relate directly to volume--average evolution,
we use the \Wv\ method -- which has the best likelihood overall -- to provide
our best estimates on the parameters $\etBg$ and
$n\ns{s}$, and the \AH\ method to estimate their
systematic uncertainties. This gives $10^{10}\etBg=6.08$ ($\pm1.5$\% stat)
($\pm8.5$\% sys) and $n\ns{s}=0.96$ ($\pm0.9$\% stat) ($\pm4.3$\% sys).
These best fit values are close to their \LCDM\ counterparts \cite{Pparm},
but the systematic uncertainties more fully reflect the limitations of our
procedure.

With these parameters the timescape cosmology remains competitive with the
\LCDM\ cosmology, also when considering other cosmological tests, as summarized
recently in Ref.\ \cite{bscg}. However, since many best fit parameters in
Table~\ref{tb:TS_params_val_Neff_fixed} are very close to \LCDM\ values, this
also means that parameter tensions may in some cases be similar. For example, a
recent determination of the radial and angular baryon acoustic oscillation
scales in the Lyman--$\alpha$ forest at a redshift $z=2.34$ found a 2.5$\,\si$
tension \cite{boss14} with the \LCDM\ parameters from Planck \cite{Pparm}.
For the timescape model the values of $\dA(z)/r\ns{drag}$ and $c/[H(z)r\ns{drag}
]$ are so close to the \LCDM\ values at $z=2.34$ \cite{boss14} that it must also
suffer tension of a similar magnitude. (However, one should add the caveat
that the treatment of redshift space distortions may need to be revisited in
non--FLRW models with backreaction \cite{rbof14}.)

Our results are of course also limited by the fact that the late time ISW
effect has not yet been computed for the timescape scenario, as it requires
a from--first--principles reanalysis. This effect is important for the
anisotropy spectrum at large angles, and we have therefore limited our
analysis to large angle multipoles $\ell>50$. If the timescape scenario is
correct, then we might also expect a nonkinematic contribution to the CMB
dipole below the scale of statistical homogeneity. Evidence for this was
found in a recent model--independent analysis \cite{hvar} of the variation
of the Hubble expansion on $\lsim120\hm$ scales. This suggests that
a detailed analysis of the large angle anisotropies, and of potential
anomalies, requires not only a computation of the late time ISW effect
in the timescape scenario, but also the impact that the subtraction of
a nonkinematic dipole component would have on the map--making procedures
in the CMB analysis.

To make more accurate predictions in the timescape scenario, it
is of course necessary to eliminate the large systematic uncertainties that
arise from the imprecise nature of the matched FLRW model procedures. This
would require a huge computational effort. However, our results highlight the
fact that important details of the acoustic peaks are strongly constrained
by the ratio of baryonic to nonbaryonic matter in the primordial plasma,
which can in principle differ in the timescape scenario.
To treat this question rigorously we must address the effect of backreaction
in the primordial plasma, and the manner in which the growth of structure is
changed when the average evolution is very close to, but not exactly, a
perturbative FLRW model. This question -- which could change some of the
conclusions of this paper -- has not been studied, but is crucially important,
not only for the timescape model but for all approaches to inhomogeneous
cosmology with backreaction.

\begin{acknowledgments}
This work was supported by the Marsden Fund of the Royal Society of New Zealand.
We have benefited from use of the publicly available computer codes \cite{lic}
CLASS \cite{class_I,class_II}, {\tt CosmoMc} \cite{Lewis_&_Bridle}, {\tt emcee}
\cite{Foreman}, {\tt fastbbn} \cite{fbbn1,fbbn2}, {\tt HyRec} \cite{HyRec}
and RECFAST \cite{RECFAST_1,RECFAST_2}. We thank Julien Lesgourgues for advice in implementing CLASS.
\end{acknowledgments}
\vskip\baselineskip
\appendix
\section{Additional relativistic species}\label{addrad}

%-------------------------------------------------%
\begin{table*}[!t]%htbp
\centering
\caption{\small\label{tb:TS_params_val_Neff_free}
The analysis of Table~\ref{tb:TS_params_val_Neff_fixed} is repeated with
$\Neff$ free to vary.}
\makebox[\textwidth]{
\begin{tabular}{ @{}l c c c c c c }
\noalign{\vskip 3pt \hrule height 1.pt \vskip 3pt }
\multicolumn{1}{@{}l}{Matching type} &
\multicolumn{2}{l}{\hbox to50mm{\hfil\AH\hfil}%FLRW $k\neq 0$, $\Lambda>0$
} &
\multicolumn{2}{l}{\hbox to50mm{\hfil\Ww\hfil}%FLRW $k = 0$, $\Lambda =0$
} &
\multicolumn{2}{l}{\hbox to50mm{\hfil\Wv\hfil}%FLRW $k \neq 0$, $\Lambda =0$
} \\
\noalign{\vskip 3pt \hrule height 0.2pt\vskip 3pt }
Parameter &\hfil Best fit\hfil & Mean (68\% limits) \hfil &\hfil Best fit\hfil & Mean (68\% limits)\hfil& \hfil Best fit\hfil & Mean (68\% limits)\hfil\\
\noalign{\vskip 3pt\hrule height 0.2pt\vskip 3pt}
$f_{\mathrm{v0}}$ & $0.637$ & $0.630 \pm 0.014$ & $0.551$ & $0.549^{+0.021}_{-0.018}$ & $0.560$ & $0.562^{+0.019}_{-0.018}$\\[5pt]

$H_0$ & $62.263$ & $61.451^{+1.047}_{-1.034}$ & $56.533$ & $56.398^{+2.162}_{-2.110}$ & $57.157$ & $57.419^{+2.395}_{-1.831}$\\[5pt]

$N_{\mathrm{eff}}$ & $3.243$ & $3.110^{+0.098}_{-0.110}$ & $3.062$ & $3.060^{+0.387}_{-0.371}$ & $3.129$ & $3.168^{+0.404}_{-0.348}$\\[5pt]

$10^{10}\eta_{\mathrm{B}\gamma}$ & $6.533$ & $6.518^{+0.108}_{-0.109}$ & $6.066$ & $6.047 \pm 0.109$ & $6.124$ & $6.112^{+0.111}_{-0.108}$\\[5pt]

$n_{\mathrm{s}}$ & $1.001$ & $0.996^{+0.011}_{-0.012}$ & $0.959$ & $0.956^{+0.020}_{-0.018}$ & $0.962$ & $0.965^{+0.020}_{-0.016}$\\

\noalign{\vskip 3pt\hrule height 0.2pt\vskip 3pt}
$\bH_0$ & $52.425$ & $51.942^{+0.513}_{-0.512}$ & $50.001$ & $49.918^{+1.487}_{-1.426}$ & $50.317$ & $50.466^{+1.598}_{-1.253}$\\[5pt]

$\Tb_0$ & $2.066$ & $2.072 \pm 0.011$ & $2.136$ & $2.138^{+0.015}_{-0.018}$ & $2.129$ & $2.127^{+0.014}_{-0.017}$\\[5pt]

$t_0$ & $16.398$ & $16.510^{+0.117}_{-0.116}$ & $16.633$ & $16.657^{+0.369}_{-0.425}$ & $16.584$ & $16.560^{+0.339}_{-0.430}$\\[5pt]

$\bar{\gamma}_0$ & $1.319$ & $1.316 \pm 0.007$ & $1.276$ & $1.275^{+0.011}_{-0.009}$ & $1.280$ & $1.282 \pm 0.009$\\[5pt]

$Y_{\mathrm{p}}$ & $0.2507$ & $0.2489^{+0.00142}_{-0.00152}$ & $0.248$ & $0.247^{+0.006}_{-0.005}$ & $0.249$ & $0.249^{+0.006}_{-0.004}$\\[5pt]

$\OMBn$ & $0.038$ & $0.039 \pm 0.001$ & $0.043$ & $0.043^{+0.002}_{-0.003}$ & $0.042$ & $0.042^{+0.002}_{-0.003}$\\[5pt]

$\OMCn$ & $0.170$ & $0.175^{+0.009}_{-0.010}$ & $0.233$ & $0.234^{+0.012}_{-0.016}$ & $0.226$ & $0.225^{+0.012}_{-0.014}$\\[5pt]

$\OMMn$ & $0.208$ & $0.213 \pm 0.010$ & $0.275$ & $0.277^{+0.014}_{-0.018}$ & $0.268$ & $0.266^{+0.014}_{-0.017}$\\[5pt]

$\OMkn$ & $0.825$ & $0.820^{+0.010}_{-0.009}$ & $0.763$ & $0.761^{+0.017}_{-0.013}$ & $0.769$ & $0.771^{+0.016}_{-0.013}$\\[5pt]

$\OM_{{\cal Q}0}$ & $-0.033$ & $-0.034 \pm 0.001$ & $-0.038$ & $-0.038 \pm 0.001$ & $-0.038$ & $-0.038 \pm 0.001$\\[5pt]

$\bar{z}_{\mathrm{dec}}$ & $1434.924$ & $1431.143^{+7.060}_{-6.898}$ & $1391.235$ & $1390.203^{+11.336}_{-9.740}$ & $1395.612$ & $1396.985^{+10.853}_{-9.437}$\\[5pt]

$\bar{z}_{\mathrm{drag}}$ & $1402.388$ & $1398.465^{+8.044}_{-7.965}$ & $1352.757$ & $1351.460^{+12.537}_{-10.983}$ & $1357.923$ & $1359.065^{+12.386}_{-10.305}$\\[5pt]

$\bD_{s}(\bar{z}_{\mathrm{dec}})$ & $0.134$ & $0.134 \pm 0.001$ & $0.132$ & $0.132 \pm 0.003$ & $0.132$ & $0.132 \pm 0.003$\\[5pt]

$\bD_{s}(\bar{z}_{\mathrm{drag}})$ & $0.139$ & $0.140 \pm 0.001$ & $0.139$ & $0.139 \pm 0.003$ & $0.138$ & $0.138^{+0.003}_{-0.004}$\\
\noalign{\vskip 3pt\hrule height 0.2pt\vskip 3pt}
$\tau_{\mathrm{w0}}$ & $13.606$ & $13.732 \pm 0.144$ & $14.228$ & $14.260^{+0.369}_{-0.449}$ & $14.145$ & $14.114^{+0.327}_{-0.448}$\\[5pt]

$\OmBn$ & $0.087$ & $0.088 \pm 0.001$ & $0.089$ & $0.089^{+0.003}_{-0.005}$ & $0.088$ & $0.088^{+0.003}_{-0.004}$\\[5pt]

$\OmCn$ & $0.391$ & $0.397 \pm 0.015$ & $0.483$ & $0.485^{+0.016}_{-0.019}$ & $0.474$ & $0.472^{+0.017}_{-0.018}$\\[5pt]

$\OmMn$ & $0.478$ & $0.486^{+0.015}_{-0.016}$ & $0.572$ & $0.574^{+0.019}_{-0.022}$ & $0.563$ & $0.560^{+0.019}_{-0.020}$\\[5pt]

$z_{\mathrm{dec}}$ & $1087.557$ & $1087.479^{+0.484}_{-0.499}$ & $1090.001$ & $1090.064^{+0.530}_{-0.535}$ & $1089.781$ & $1089.869^{+0.563}_{-0.551}$\\[5pt]

$z_{\mathrm{drag}}$ & $1062.892$ & $1062.638^{+0.831}_{-0.839}$ & $1059.849$ & $1059.672^{+1.401}_{-1.412}$ & $1060.345$ & $1060.272^{+1.520}_{-1.263}$\\[5pt]

$100\,\theta_{\mathrm{dec}}$ & $1.047$ & $1.047 \pm 0.001$ & $1.041$ & $1.041 \pm 0.001$ & $1.041$ & $1.041 \pm 0.001$\\[5pt]

$100\,\theta_{\mathrm{drag}}$ & $1.063$ & $1.064 \pm 0.001$ & $1.061$ & $1.061 \pm 0.002$ & $1.060$ & $1.060^{+0.001}_{-0.002}$\\[5pt]

$d_{\mathrm{A}}$ & $12.774$ & $12.840 \pm 0.101$ & $12.706$ & $12.724^{+0.279}_{-0.283}$ & $12.686$ & $12.673^{+0.249}_{-0.310}$\\[5pt]

$d_{\mathrm{A,drag}}$ & $13.065$ & $13.136^{+0.097}_{-0.096}$ & $13.062$ & $13.083^{+0.296}_{-0.304}$ & $13.033$ & $13.022^{+0.264}_{-0.331}$\\
\noalign{\vskip 3pt \hrule height 1.pt \vskip 3pt}
\end{tabular}}
\end{table*}
%-------------------------------------------------%
We have also investigated the case when $\Neff$ is left free to vary as an
additional MCMC parameter, without elaborating what the extra radiation
component represents. We do not aim to constrain neutrino masses, for example.

Varying $\Neff$ affects the Boltzmann equations directly, and also the BBN
helium abundance, $Y_{\rm p}=Y_{\rm p}(\Neff,\etBg)$, which is passed to
other parts of the code.

We find that the results with $\Neff$ free to vary do not differ significantly
from the case $\Neff=3.046$ of the base model. The results are shown in Table
\ref{tb:TS_params_val_Neff_free} for the canonical \AH\ matching method. One
reason for this investigation was to check whether the lithium anomaly problem
could be alleviated at all by leaving $\Neff$ free to deviate from $3.046$.
Deviations are possible if there are extra radiation components or a
neutrino/antineutrino asymmetry, to list a couple of scenarios. This could be
possible if BBN determination of $Y_{\rm p}=Y_{\rm p}(\Neff,\etBg)$ forced
$\etBg$ towards lower values, $\etBg\approx (5.1\pm0.5)\times10^{-10}$, which
resolve the lithium abundance anomaly \cite{bbn,pill}.

On the contrary we found preference for $\Neff>3.046$ and because $\Neff$ is
positively correlated with $\etBg$ this leads to an even slightly larger
$\etBg$. This is accompanied with a preference for increased $\fvn$ because
$\etBg$ is positively correlated with $\fvn$ but negatively correlated with
$\OmMn$. (See Fig.~\ref{fig:probs_chains_ts_lcdm_rect_dressed}.)\vfil

\end{document}